\documentclass[universe,article,pdftex,moreauthors]{Definitions/mdpi} 
\preto{\abstractkeywords}{\nolinenumbers}
\firstpage{1} 
\pubvolume{1}
\issuenum{1}
\articlenumber{0}
\pubyear{2023}
\copyrightyear{2023}
\datereceived{ } 
\daterevised{ } 
\dateaccepted{ } 
\datepublished{ } 
\hreflink{https://doi.org/} 

\usepackage{mathtools}
\usepackage{gensymb}
\usepackage{tikz}
\newcommand{\tikzxmark}{%
\tikz[scale=0.23] {
    \draw[line width=0.7,line cap=round] (0,0) to [bend left=6] (1,1);
    \draw[line width=0.7,line cap=round] (0.2,0.95) to [bend right=3] (0.8,0.05);
}}

\usepackage{graphicx}	
\usepackage{amsmath}	
\usepackage{amssymb}	

\newcommand{\myarrow}{\tikz\draw[thin,black,-latex] (1,4.5ex) -- ++(0,-2.0 5ex) -- +(4.5ex,0); \;}

\usepackage[normalem]{ulem}

\Title{Energetic particles and high-energy processes in cosmological  filaments 
  and their astronomical implications}

\TitleCitation{Energetic particles and high-energy processes in cosmological filaments}

\Author{Kinwah Wu $^{1,4}$*\orcidA{}, Ellis R. Owen $^{2}$*\orcidB{}, 
Qin Han$^{1}$\orcidC{},  
Yoshiyuki Inoue$^{2, 3, 4}$\orcidD{}, 
  Lilian Luo $^{1}$\orcidE{}
 }

\AuthorNames{Kinwah Wu, Ellis R. Owen, 
Qin Han, Yoshiyuki Inoue and Lilian Luo}

\AuthorCitation{Wu, K.; Owen, E. R.; Han, Q.; 
 Inoue, Y.; Luo, L.}

\address{%
$^{1}$ \quad  Mullard Space Science Laboratory, University College London, Holmbury St.~Mary, Surrey RH5 6NT, UK\\
$^{2}$ \quad Theoretical Astrophysics, Department of Earth and Space Science, Graduate School of Science, Osaka University, Toyonaka 560-0043, Osaka, Japan  \\ 
$^{3}$ \quad Interdisciplinary Theoretical \& Mathematical Science Program (iTHEMS), RIKEN, 2-1 Hirosawa, Saitama 351-0198, Japan \\
$^{4}$ \quad Kavli Institute for the Physics and Mathematics of the Universe (WPI), UTIAS, The University of Tokyo, Kashiwa, Chiba 277-8583, Japan
}

\corres{Correspondence: 
kinwah.wu@ucl.ac.uk (K.W.), 
erowen@astro-osaka.jp (E.R.O.)}

\abstract{Large-scale cosmic filaments 
  connect galaxies, clusters and voids. 
 They are permeated by magnetic fields with a variety of topologies. 
Cosmic rays with energies up to $10^{20}\;\!{\rm eV}$ can be 
  produced in astrophysical environments 
  associated with star-formation 
  and AGN activities. 
The fate of these cosmic rays in filaments, 
  which cannot be directly observed 
  on Earth, are rarely studied. 
We investigate 
  the high-energy processes 
  associated with energetic particles 
  (cosmic rays)  
  in filaments,  
  adopting an ecological approach 
  that includes  
  galaxies, clusters/superclusters 
  and voids 
  as key cosmological structures  
  in the filament ecosystem.  
We derive the phenomenology 
  for modelling interfaces  
  between filaments and these structures,   
  and investigate 
  how the transfer and fate of 
  energetic cosmic ray protons 
  are affected 
  by the magnetism of the interfaces.   
We consider different magnetic field 
 configurations in filaments 
 and assess the implications for 
 cosmic ray confinement and survival against hadronic pion-producing and photo-pair interactions. 
Our analysis shows that the fate of the particles 
 depends on the location of their origin 
 within a filament ecosystem, and that 
 filaments act as `highways', channelling cosmic rays 
  between galaxies, 
 galaxy clusters and superclusters. 
Filaments can also operate as cosmic `fly paper', 
  capturing cosmic ray protons   
  with energies up to $10^{18}\;\!{\rm eV}$ 
  from cosmic voids. 
Our analysis predicts the presence of  
  a population of $\sim 10^{12}-10^{16}\;\!{\rm eV}$ 
  cosmic ray protons in filaments and voids 
  accumulated continually over cosmic time. 
These protons  
  do not suffer significant energy losses  
  through photo-pair or pion-production,    
  nor can they be cooled 
  efficiently.  
Instead, they form a cosmic ray fossil record 
  of the power generation history of the Universe. }


\begin{document}




%
\section{Introduction}

Filaments, walls and voids 
  are the largest structures in the Universe. 
Filaments are diffuse media 
  connecting lower-order hierarchical structures,  
   such as the gravitationally-bound  
   galaxy clusters and groups. 
They contain dark matter and baryons (gas), 
  but the baryons contribute 
  only about $5-10$\% of the filament mass 
  \citep[see e.g.][]{Eckert2015Nat,Vernstrom2021MNRAS}. 
The baryon density in filaments 
  is about $10 - 100$ times 
  higher than the cosmic average value 
  at redshift $z=0$ \citep{Tanimura2020A&A,Reiprich2021A&A,Vernstrom2021MNRAS}, 
   which is about $4 \times 10^{-31}{\rm g}~{\rm cm}^{-3}$   
  \citep{Walter2020ApJ}. 
This gas is considered as a candidate for a major invisible ``missing'' constituent of baryons in the Universe (cf. the ``missing baryon problem''; Ref.~\cite{Nicastro2018Nat}). 
Baryons in filaments mostly consist of 
 hot ionised gas 
  and warm partially ionised gas. 
This multi-phase mixture is 
  often referred to as the warm-hot intergalactic medium (WHIM).
The current view is that WHIM 
  is heated by shocks 
  generated by energetic events 
  associated with structure formation, together with feedback processes from galaxies, 
  e.g. galaxy mergers, galactic outflows 
  or active galactic nuclei (AGN) activities \citep[see][]{Nath2001MNRAS,Scannapieco2005ApJ,Cen2006ApJ}. 

The Universe is permeated by magnetic fields, 
  and naturally filaments are magnetised.   
The strengths of their magnetic fields 
  inferred from recent observations 
  is $\sim 10 - 60\;\!{\rm nG}$  \citep{Vacca2018Galax,Vernstrom2021MNRAS,Carretti2023MNRAS}, but their origin is yet to be resolved.   
While they could be partly cosmological in nature 
  \citep[see][]{Carretti2023MNRAS}, 
  star-forming activities in galaxies 
  may play a non-negligible role in magnetizing filaments 
\citep{Kronberg1994RPPh,Bertone2006MNRAS,Vazza2017CQGra}, 
  especially in regions 
  where the filaments meet 
  the circumgalactic medium (CGM) 
  \citep[see][]{Heesen2023A&A}. 

Stacking observations in X-rays and radio wavebands 
  \citep[see][]{Akamatsu2017A&A,Vernstrom2021MNRAS} 
  indicate the presence of highly energetic particles,  
  often referred to as cosmic rays, in filaments.  
(Here and hereafter, unless otherwise stated, 
  we adopt the terminology 
  energetic particles and cosmic rays 
 interchangeably.)   
There is no direct observational evidence that 
  the cosmic rays and magnetic fields in filaments 
  are in energy equipartition.  
The cosmic ray particles may be accelerated in situ 
  in the filaments, 
  e.g. through accretion shocks near structures 
   \citep{Vernstrom2023SciA}, 
  or they may be produced elsewhere,   
  e.g. in star-forming galaxies 
  \citep[see e.g.][]{Thomas2023MNRAS} 
  or in AGN and their jets 
  \citep[see e.g.][]{Ptuskin2013AdSpR}, 
  then transported into the filaments.  
The structure and thermodynamics of filaments 
   evolve as consequence of cosmological structural 
   formation dynamics 
   \citep[see][]{Galarraga-Espinosa2023arXiv}. 
Star-forming and AGN activities 
  are also not uniform across cosmic history.  
Hence,  
  the properties and composition 
  of energetic particles in filaments 
  evolve over time.  
  
In this work,  
  we investigate the high-energy processes 
  associated with energetic particles 
  (cosmic rays) 
  in cosmological filaments, 
  and determine 
  their consequences. 
We organise the paper as follows. 
In Sec.~2, we 
    describe hadronic processes in 
    astrophysical environments; 
in Sec.~3, we  
  elaborate how energetic particles 
  are magnetically confined 
  in various configurations 
  of filament magnetic fields;    
 and in Sec.~4, we illustrate how particles 
  are transferred between 
  filaments   
  and voids, superclusters/clusters and galaxies by following the cosmic journey of
  individual particles.  
In Sec.~5, we discuss  
  the implications for  
  the presence 
  of populations of energetic particles 
   in cosmic filaments     
   resulting from the  
   interactions and transfer processes 
   in filament ecosystems.  
A short conclusion is presented in Sec.~6.   

%
\section{Hadronic interactions 
  in astrophysical environments} 
\label{sec:hadronic_interactions}

Cosmic rays are 
  a mix of particles of different species.  
At energies below $10^{19}\;\!{\rm eV}$, 
  hadronic cosmic rays 
  in galactic and extragalactic environments  
   are believed to be mostly protons (H nuclei). 
Heavier nuclei dominate at higher energies. 
This seems to be supported 
  by the observed composition of cosmic rays 
  arriving on Earth 
  (see, e.g. analysis of data obtained 
  by the Pierre Auger Observatory 
  \cite[][]{Kampert2014CRPhy}). 
The dominance of heavy nuclei at energies 
   above $10^{19}\;\!{\rm eV}$ 
  implies that the transport of multi-species cosmic rays 
  and the acceleration of heavy nuclei 
  are more complicated 
   than scenarios of cosmic ray attenuation 
  based solely on the Greisen-Zatespin-Kuzmin (GZK) effect \citep[][]{Kampert2017AIPC,Owen2021ApJ}.   
Adding to this complexity,
  the composition of cosmic rays 
  and their properties are
  not uniform in space or  
  over cosmological time.  
It is a challenging task to disentangle 
  these factors and 
  the effects they induce,  
  given that our understanding of structural formation 
  at the sub-cluster and galactic levels 
  is still incomplete 
  and our capability to confidently 
  identify and model cosmic particle accelerators 
  beyond the framework 
  of stochastic processes in shocks 
  is limited.  
Even on the scale of the Solar system, 
  we cannot rely on information 
  about the composition of cosmic rays 
  arriving on Earth to infer the energy distribution and composition of cosmic rays 
  in local interplanetary space.  
For example, 
  cosmic ray baryons observed on Earth at sea level are, 
  in fact, mostly neutrons 
  \citep[][]{Ziegler1998IBMJ,Sato2015PLoSO}.  
Some in situ measurements by spacecraft 
  have extended the domain of our direct measurements 
  of cosmic rays \citep[][]{Stone2019NatAs}, 
  extending to 
  the edge of the Solar system 
  \citep[][]{Zhang2015PhPl,Rankin2019ApJ}
  and local interstellar space 
  beyond the heliosphere 
  \citep{Cummings2016ApJ}.   
However, beyond the Solar system 
  and immediate local region of interstellar space, 
  our knowledge of cosmic rays 
  can only be inferred from simulations  
  \citep[see][]{Kissmann2014APh,Werner2015APh},   
  or phenomenological modelling 
  \citep[e.g.][]{Globus2015PhRvD,Kempski2022MNRAS, Ambrosone2022MNRAS, Phan2023PhRvD}, 
   often based on information 
  derived from observations in $\gamma$-rays  
  \citep[e.g.][]{Aharonian2020PhRvD,Ajello2020ApJ, Tibaldo2021Univ}, 
  radio \citep[e.g.][]{McCheyne2022A&A,Yusef-Zadeh2024MNRAS}, 
  or at other wavelengths 
  \citep[e.g.][]{Indriolo2018ApJ, Okon2020PASJ, Bialy2020CmPhy,Pineda2024arXiv}.  
To date, volumes of manuscripts have been published 
  in efforts devoted towards understanding 
  cosmic ray composition and their properties 
  in galaxies 
  and in intergalactic media (IGM). 
Despite this, it remains a subject 
  of ongoing discussion and debate 
  \citep[for recent reviews, see][]{Kotera2011ARA&A, Grenier2015ARA&A,Ruszkowski2023A&ARv,Owen2023Galax}. 


Cosmic rays interact with radiation and baryons 
 in interstellar and intergalactic space,  
 hence their content and composition 
  evolve  
  as they propagate. 
Without loss of generality, 
  we illustrate the interactions   
  with a proton (or a neutron) interacting 
  with a photon or with another proton. 
These processes are referred 
  to as p$\gamma$ 
  (or n$\gamma$ for neutron) and pp interactions,  
  respectively. 
The p$\gamma$ interaction is dominated by two channels. The first is photo-pion production. This,  
 including the subsequent interactions 
 and its decay branching, 
proceeds as follows (resonant states not shown) 
\citep[see e.g.][]{Dermer2009herb}:  
\begin{align}%
\label{eq:pg_int}%
 {\rm p}+  \gamma \longrightarrow 
	\begin{cases}%
	  \;  {\rm p} \pi^0 \longrightarrow {\rm p}\;  2\gamma				\\[0.5ex]%
	  \;  {\rm n} \pi^+ \longrightarrow {\rm n}\;\! \upmu^+ \nu_{\upmu}		\\%
		\hspace{5.5em} \myarrow {\rm e}^+ \nu_{\rm e} \bar{\nu}_{\upmu}%
	\end{cases}  \ ;   \\ 
%
 {\rm n} + \gamma \longrightarrow 
		\begin{cases} %
	\; 	{\rm p}\;\! \pi^-  \longrightarrow 
    {\rm p}\;\! \upmu^- \bar{\nu}_{\upmu}		\\ %
		\hspace{5.5em} \myarrow {\rm e}^- {\bar\nu}_{\rm e} {\nu}_{\upmu}%
    \\  
		\; 	{\rm n}\;\! \pi^0 \longrightarrow {\rm n}\;  2\gamma	  
		\end{cases} \ .  %
\end{align}%
The minimum energy for a proton, 
 $E_{\rm p}\;\! (= \gamma m_{\rm p} [c]^2)$, 
  to initiate 
  a chain of pion production 
  in a radiation field is given by  
\begin{align} 
  \gamma_{\rm p} = \frac{1}{4}\;\! \frac{m_{\rm p}[c]^2}{\epsilon_{\rm ph}} 
  \left[ \left( \frac{m_{\rm n}}{m_{\rm p}} 
  + \frac{m_{\pi^+}}{m_{\rm p}} \right)^2 - 1 \;\! \right]  
    \approx \frac{2}{25}  \;\!  
    \frac{m_{\rm p}[c]^2}{\epsilon_{\rm ph}}  \      
\label{eq:proton_energy_01}
\end{align} 
 (for $m_{\rm p}[c]^2 \gg \epsilon_{\rm ph}$), 
 where $c$ is the speed of light, 
 $\gamma_{\rm p}$ is the Lorentz factor of the proton, 
 $\epsilon_{\rm ph}$ is the energy of the photon, and 
 $m_{\rm p}$, $m_{\rm n}$ and $m_{\pi^+}$ 
 are the masses of the proton, neutron and charged pion, respectively.  

 The second channel of the p$\gamma$ interaction is Bethe-Heitler photo-pair production \cite{Bethe1934RSPSA}. 
 This proceeds as follows: 
 \begin{align}
     {\rm p}' + \gamma \rightarrow {\rm p} + l^+ + l^- \ ,   
 \end{align}
 where ${\rm p}'$ and ${\rm p}$ are the cosmic ray protons before and after the pair-production process, respectively. $l^{\pm}$ are the produced lepton/anti-lepton pair, which are dominated by electron and positrons~\cite{Blumenthal1970PhRvD, Klein2006RaPC} 
 although heavier leptons can also be formed~\cite{Hooper2023PhRvL}. 

The major channels for the pp interaction
  and their branching 
  are as follows:   
\begin{align}
\label{eq:pp_interaction} 
{\rm p} + {\rm p} \longrightarrow 
	\begin{cases} 
	\; 	({\rm p}\;\!  \Delta^{+~} )\;\! \ \longrightarrow  \begin{cases} 
				{\rm p}\;\! {\rm p}\;\! \pi^{0} \;\! \xi_{0}(\pi^{0})\;\! \xi_{\pm}(\pi^{+} \pi^{-}) \\[0.5ex] 
				{\rm p}\;\! {\rm p}\;\!  \pi^{+}  \pi^{-}\;\!  \xi_{0}(\pi^{0})\;\! \xi_{\pm}(\pi^{+} \pi^{-}) \\[0.5ex] 
				{\rm p}\;\! {\rm n}\;\!  \pi^{+}\;\!  \xi_{0}(\pi^{0})\;\! \xi_{\pm}(\pi^{+} \pi^{-})\\[0.5ex] 
			\end{cases} \\ 
	\; (	{\rm n}\;\! \Delta^{++} ) \longrightarrow  
            \begin{cases}
				{\rm n}\;\! {\rm p}\;\! \pi^{+}\;\! \xi_{0}(\pi^{0})\;\! \xi_{\pm}(\pi^{+} \pi^{-}) \\[0.5ex]
				{\rm n}\;\! {\rm n}\;\! \pi^+ \pi^{+} \xi_{0}(\pi^{0})\;\! \xi_{\pm}(\pi^{+} \pi^{-})\\[0.5ex]
			\end{cases} \\
	\end{cases} \ . 
\end{align}%
Unlike the p$\gamma$ interaction, 
  the dominant channels of the pp interaction 
  tend to produce resonance particles \citep[e.g.][]{Almeida1968PhRv, Skorodko2008EPJA, Kafexhiu2014PhRvD}, 
  such as $\Delta^+$ and $\Delta^{++}$. 
  Their subsequent decays give rise to multiple pions. 
  Among these,  
   neutral pions produce $\gamma$-rays  
  while charged pions produce leptons 
  and their corresponding neutrinos. 
In pp interactions,  
   decays of the $\Delta$ resonances  
  restore the number of strange-zero ($S=0$) baryons 
  of the $S=1/2$ ground-state baryon octet 
  while producing the lowest-mass strange-zero pseudo-scalar mesons 
    of the spin-zero nonet. 

In the centre-of-momentum frame, 
  the threshold energy of the protons, ${\tilde \gamma}_{\rm p}m_{\rm p}[c]^2$, 
  for pion-production in a pp interaction 
  is the available energy of the protons in a collision 
  which excites a p$\Delta^+$ or a n$\Delta^{++}$ intermediate state: 
\begin{align} 
\begin{cases} 
\ 2\;\! {\tilde \gamma}_{\rm p} m_{\rm p} \approx m_{\rm p} +m_{\Delta^+} \\ 
\ 2\;\! {\tilde \gamma}_{\rm p} m_{\rm n} \approx m_{\rm p} +m_{\Delta^{++}} 
\end{cases} \ , 
\end{align} 
which gives  
\begin{align}
  {\tilde \gamma}_{\rm p} \approx \frac{1}{2}\ \left[\;\! \frac{m_{\rm x}}{m_{\rm p}} 
  + \frac{m_\Delta}{m_{\rm p}} \;\! \right] \ , 
\end{align} 
  where ${\rm x} \in \{{\rm p},{\rm n}\}$ 
  and $m_\Delta$ is the mass of the $\Delta$ resonance particle, 
  which is about $1.232~{\rm GeV}/[c]^2$.   
The pions resulting from the decay 
  of the $\Delta$ particles 
  therefore 
  can not be at rest 
  in the centre-of-momentum frame, 
  but instead have a substantial amount of kinetic energy. 
Moreover, 
  there would be a dichotomy in the energy distribution of the pions. 
Thus, the pions retain certain information 
  about the energetic protons 
  that initiate a pp interaction.  

The proton threshold energy 
   for pion-production  
   in a pp interaction 
  is only slightly above $m_{\rm p}[c]^2$, 
  and is insensitive to parameters 
  other than the rest masses of the particles involved. 
The situation is very different 
 in the p$\gamma$ interaction,  
 where the threshold energy of the protons 
  is dependent on the photon energy 
  in the radiation field. 
The wavelength of the CMB 
  (cosmic microwave background radiation) 
  in the current epoch 
  (redshift $z =0$) is about 2~mm.  
The CMB spectrum 
    has a peak photon energy 
    of $\epsilon_{\rm ph} \approx 6.63 \times 10^{-4}~{\rm eV}$.  
The proton threshold energy is therefore   
  $E_{\rm p} \approx  10^{20}\;\!{\rm eV}$   
  (for $\gamma_{\rm p} > 1.1 \times 10^{11}$).  
Starburst galaxies tend to have  
  a prominent infra-red (IR) emission component,
  peaking at wavelengths   
  $\sim 50 - 100\;\! {\mu}{\rm m}$ 
  \citep[e.g. M82, see][]{Galliano2004PhDT,Schreiber2018AA}. 
The photon energy at $75\;\!{\mu}{\rm m}$   
  is $\epsilon_{\rm ph} \approx 1.65\times 10^{-2}\;\!{\rm eV}$,  
  giving a proton threshold energy of 
$E_{\rm p} \approx 4.3 \times 10^{18}\;\!{\rm eV}$ 
  (for $\gamma_{\rm p}> 4.6 \times 10^9$).   
The spectrum of a disk or elliptical galaxy 
  generally peaks  
  at wavelengths around $1.0\;\!{\mu}{\rm m}$ 
  \citep[e.g. M101 and NGC 5018, see][]{Galliano2004PhDT}, 
  and a substantial fraction of the photons 
  in the radiation field would have energies of 
  $\epsilon_{\rm ph} \approx 1.25\;\!{\rm eV}$, 
  which implies a characteristic proton threshold energy of $E_{\rm p} \approx 5.6 \times 10^{16}\;\!{\rm eV}$ 
  (for $\gamma_{\rm p}> 6.0 \times 10^7$). 
  




\section{Confinement and trapping of energetic particles}   
\label{sec:confinement} 

\subsection{Filaments as mass condensates 
  and particle interactions} 
\label{subsec:filaments}

Because of the lack of direct observations,  
  our current knowledge of large-scale cosmic filaments is primitive.  
Not much is known about their geometrical properties  
    (such as their thickness), 
their dark matter distribution, the thermodynamic properties of filament gas,
the configurations and origin(s) 
     of filament magnetic fields,
or their evolution over cosmological time. 
Our understanding of cosmic filaments 
  is derived mostly from 
  numerical simulations \citep[see][]{Aragon-Calvo2010MNRAS,Cautun2014MNRAS,Zhu2021ApJ,Gouin2022A&A}.  
Despite this,  
  there is no doubt that 
  filaments are very important components 
  in the hierarchy of cosmological structures. 
Indeed, most of the mass in the Universe 
  is confined in filaments in the current epoch, slightly exceeding the total mass contained in galaxy clusters
  \cite{Aragon-Calvo2010MNRAS}. 

Clusters are young objects. 
While proto-clusters with redshifts 
  as high as $z\sim (6-7)$ are present 
  \citep[][]{Harikane2019ApJ}, and some clusters with a developed 
  thermalised intra-cluster medium (ICM) 
  could have formed slightly above $z\sim 2$ 
  \cite[see][]{DiMascolo2023Natur}, 
  the majority of fully-fledged clusters are found to reside 
  at $z < 1$ 
  \cite[see][]{Wen2012ApJS}.  
Filamentary structures, in contrast, 
   appear well before $z\sim 4$ 
  \cite[see][]{Zhu2021ApJ}, 
  implying these have always been 
  the dominant mass condensates of the Universe.  
Embedded in high-$z$ filaments 
  are galaxies and groups of galaxies 
  in the early stages of their lives. 
Stars are formed in these galaxies,   
  and supermassive black holes 
  grow through merging and accretion.  
These black holes 
  would appear as AGN 
  when they accrete gas 
  from their surroundings.  
This has very significant implications 
  from the perspective of multi-messenger astronomy 
  involving highly energetic non-photonic particles. 

It is generally accepted that 
  starburst galaxies and AGN 
  are capable of producing energetic particles 
  of energies from a GeV to above a PeV 
  \citep{VERITAS2009Natur, Kotera2011ARA&A, Romero2018A&A, Lunardini2019JCAP}. 
Cosmic star-formation activities peak 
  at $z\sim 2$ \citep{Hopkins2006ApJ} 
  (often referred to as the ``cosmic noon''
  \citep[see][for a review]{Forster_Schreiber2020ARA&A}). 
 AGN activities also peak at $z\sim 2$ 
   \cite[][]{Wolf2003A&A}.   
This implies that 
  most of the energetic particles 
 produced in the Universe 
  must pass through filaments 
  before they can escape to the cosmic voids, 
  or are somehow trapped in filaments 
  after they have left 
  their galaxy of origin. 
Hadronic particles in cosmic voids, if they have sufficient energies, 
 interact with CMB 
  photons 
  to produce lower energy hadronic particles 
  and leptons.  
They also lose some fraction of their energy through adiabatic process  
  due to cosmological expansion.  

Like clusters and galaxies, filaments evolve.  
Simulations show that they stretch and thicken over time 
  \citep[see e.g.][]{Aragon-Calvo2010MNRAS,Cautun2014MNRAS,Zhu2021ApJ}.  
The density and thermal conditions   
  of the gas and particles in filaments  
  therefore does not stay constant.   
Filaments are in fact large-scale ecosystems.  
They exchange energy and chemicals  
  with the galaxies embedded within them, and the clusters 
  hooked onto them 
  through processes such as accretion and outflows \cite[see e.g.][]{Gray2013ApJ}. 
In addition, filaments are also irradiated 
  by the stars in galaxies and by AGN, 
  which could be a heating source that can modify their thermal conditions. 

The evolution of the thermal 
  and mechanical properties of filaments, 
  together with the development of filament magnetic fields, 
   determines whether energetic particles 
  (including nuclei, bayons and leptons) 
  produced by star-formation or AGN activities  
  can break the confinement of their host galaxies.  
Generally, energetic heavy nuclei will degrade into
  lighter nuclei or single baryons 
  through spallation collisions or hadronic interactions.  
Cosmic ray baryons will interact with photons in radiation field 
  or with other baryons, resulting 
  in lepton pair and pion-production 
  (see \S~\ref{sec:hadronic_interactions}). 
 Charged pions then decay to produce leptons and neutrinos,  
 while neutral pions decay to form $\gamma$-ray photons.
 Heavier leptons will decay into lighter leptons  
  and their corresponding neutrinos, 
  eventually to electrons/positrons  
  and electron neutrinos. 
Energetic electrons and positrons gradually
 lose their energy   
  through radiative processes.   

An important factor  
  that determines whether energetic particles in filaments 
 undergo pp and p$\gamma$ interactions 
  is the size of filaments. 
Maps of large-scale structures in the Universe  
  have shown that filaments 
  have lengths of several tens of Mpc 
  \citep{Bharadwaj2004ApJ,Sarkar2023MNRAS}, 
  with the longest ones 
  exceeding 100~Mpc \citep{Pandey2011MNRAS}. 
The thickness of filaments is not easily 
  determined directly from observations, however 
simulations have indicated that 
  the thickness of filament spines 
  are $\sim 2\;\!{\rm Mpc}$ 
  \citep[see e.g.][]{Aragon-Calvo2010MNRAS,Cautun2014MNRAS,Wang2021NatAs}, 
  and filaments in the current epoch ($z =0$) 
  could have a thickness of $\sim 4\;\!{\rm Mpc}$   
  \citep{Zhu2021ApJ}.

%
%
\begin{figure}[H]
\begin{adjustwidth}{-\extralength}{0cm}
\centering
\includegraphics[width=8.5cm]{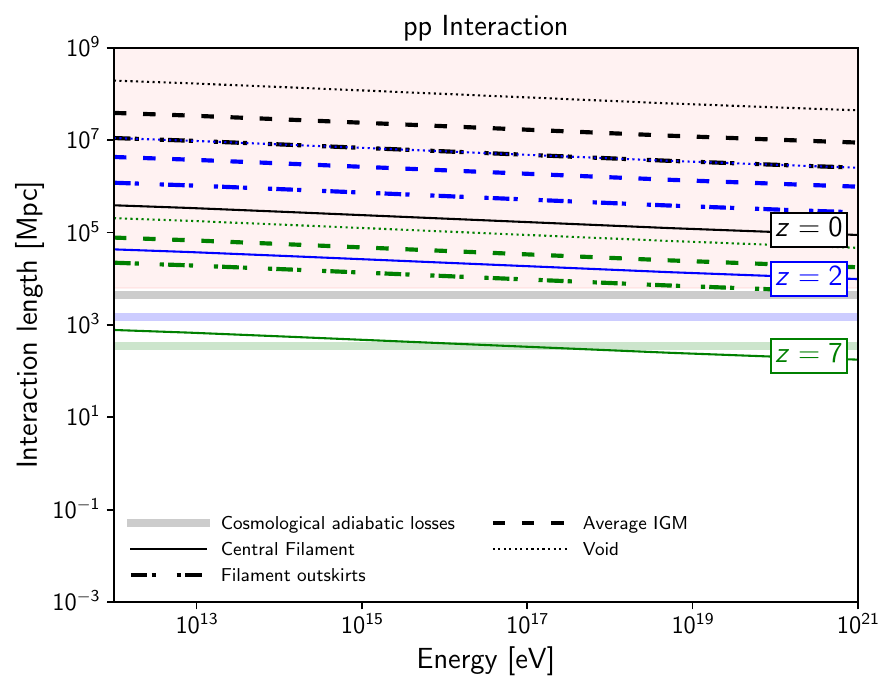} 
\includegraphics[width=8.5cm]{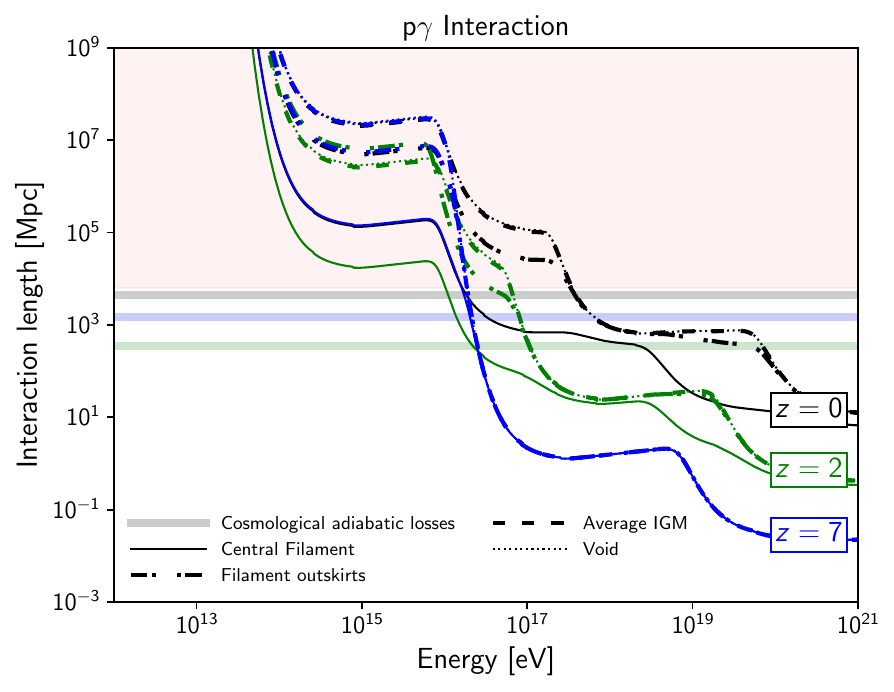} 
\end{adjustwidth}
\caption{The interaction lengths of protons undergoing 
 pp interaction processes (left panel) 
  and p$\gamma$ processes (right panel) 
  for photo-pair and photo-pion production in cosmic filaments and voids, at redshifts of $z=0$, 2 and 7. 
The calculations are based on those for the hadronic pp and p$\gamma$ interactions in Ref. \citep{Owen2019A&A}.  
The parameters adopted for the calculations 
  are shown in Appendix~\ref{app:AAAAA}. 
The length-scales 
  for proton adiabatic losses 
  at $z=0$, 2 and 7 
  due to cosmological expansion are also shown 
   for comparison. These assume a $\Lambda$CDM cosmology with cosmological parameters of $H_0 = 67.4$ km s$^{-1}$ Mpc$^{-1}$, $\Omega_{\rm m} = 0.315$, $\Omega_{\Lambda} = 0.685$ and  negligible curvature and radiation energy densities (following the 2018 \textit{Planck} results, Ref.~\cite{Planck2020A&A}). 
The distance to the event horizon of the Universe 
  at the current epoch 
  is roughly the same as the adiabatic loss 
  length-scale of protons at $z=0$. 
Interaction lengths above this scale  
  (indicated by the pink shaded regions in the panels) 
  are not of astrophysical consequence, 
   but are shown for completeness. } 
\label{fig:hadronic_ints}
\end{figure}  
%
%

Fig.~\ref{fig:hadronic_ints} shows 
  the interaction lengths 
  of pp and p$\gamma$ interactions 
  in cosmological filaments and voids
  at redshifts $z= 0$, 2 and 7 (respectively corresponding to the current epoch, the 
  cosmic noon when star-formation and AGN activities 
  peaked, and the cosmic dawn 
  during the process of cosmological reionisation). The parameters adopted for the calculations are summarised in Appendix~\ref{app:AAAAA}. 
Given that the width of filaments 
  would not exceed 4~Mpc at any epoch, 
  pp or pp-like interactions are inconsequential 
  for unconfined energetic baryons, 
  which are unaffected in filaments and voids. 
At the current epoch ($z=0$),  
  unconfined protons are not attenuated 
  by p$\gamma$ interactions in filaments;  
   only protons with energies above 
  $\sim 3 \times 10^{20}~{\rm eV}$ 
  would be degraded by interactions 
  with CMB photons  
  (cf. the GZK effect \citep[][]{Greisen1966PhRvL,Zatsepin1966JETPL}). 
At the cosmic noon ($z = 2$), 
  unconfined protons of energies below $10^{19}~{\rm eV}$  
  would not be affected when traversing 
 a filament, but 
  protons with energies above $10^{17}~{\rm eV}$ 
  could undergo p$\gamma$ interactions 
  when they travel along a filament or cross a void. 
At the cosmic dawn ($z =7$), 
  unconfined protons of energies above $10^{17}$~eV 
  would be attenuated by p$\gamma$ interactions in both filaments and voids. 

The Universe has a finite age 
  and the cosmic horizon  has a finite extent. 
Without deflections,  
  energetic protons could propagate 
 over distances of a few Gpc.  
From this, 
  together with the interaction length comparisons 
   shown in Fig.~\ref{fig:hadronic_ints}, 
  we come to the following conclusions:  
(i) cosmic ray protons with energies below about 
  $10^{16}~{\rm eV}$ will not be attenuated 
  in filaments or voids; 
(ii) energetic protons do not directly 
  deposit energy into filament gas; and 
(iii) energetic protons below $10^{16}~{\rm eV}$ 
  retain a substantial amount of their energy  
  when they are confined in filaments,  
  until they collide with other baryons 
  or high-energy photons 
  (such as the keV X-rays from AGN).

%
\subsection{Gyration of charged particles} 
\label{subsec:particle_gyration} 

The radius of gyration, the Larmor radius $r_{\rm L}$, 
  of a particle  with a charge $Ze$ and a mass $m$ 
  around a magnetic field ${\boldsymbol B}$ 
  is given by 
\begin{align}  
r_{\rm L} & = \gamma\beta   \sin \theta \  
  \frac{mc^2 }{|Ze {\boldsymbol B}|} 
   = 1.7\times 10^{12}\ \gamma \beta 
   \left(\frac{\sin \theta}{Z}\right)     
   \left(\frac{m}{m_{\rm e}}\right) 
   \left(\frac{|{\boldsymbol B}|}{\rm nG} \right)^{-1} {\rm cm}
   \ .  
\end{align}  
Here, $\gamma ( = (1- \beta^2)^{-1/2})$ 
  is the Lorentz factor of the particle, 
  ${\boldsymbol \beta}$ the velocity of the particle 
  (normalised to the speed of light $c$) 
  and $\theta (= \cos^{-1} 
({{\boldsymbol \beta} \cdot {\boldsymbol B}}/{|\boldsymbol B|}))$ is 
  its pitch angle.  
We may define a parameter 
\begin{align}   
\zeta_{\rm L} \equiv \frac{1}{2\pi{\mathcal D}} 
    \int_{2\pi} {\rm d} \Omega \ \ r_{\rm L}  \ ,  
\end{align}
  where $\mathcal D$ 
  is the characteristic size of a domain 
   in which the threading magnetic field 
    has a coherent structure. 
For a relativistic nucleus ($\beta \rightarrow 1$, $\gamma \gg 1$) 
  of mass number $A$,   
\begin{align} 
\zeta_{\rm L,nu} \approx 8.5 \times 10^{-8}\ 
  \left(\frac{A}{Z} \right) 
   \left(\frac{E_{\rm nu}}{1\;\!{\rm TeV}} \right)  
  \left(\frac{\mathcal D}{1\;\!{\rm Mpc}} \right)^{-1}  
  \left(\frac{ {\mathcal B}}{10\;\!{\rm nG}} \right)^{-1} \ ,  
\label{eq:zeta_nu}
\end{align} 
where $E_{\rm nu}$ is the energy of the nucleus. 
The parameter $\zeta_{\rm L,nu}$  
  is a measure of whether or not 
  a charged nucleon would be confined 
  in a domain of extent ${\mathcal D}$ 
  with a coherent magnetic structure 
  of a characteristic field strength 
  ${\mathcal B}$. 
This variable $\zeta_{\rm L,nu}$ has 
  no explicit dependence on the mass of the charged particle, but instead depends on the 
  ratio $(A/Z)$. The parameter $\zeta_{\rm L,nu}$ not only 
  determines if a nucleon of $A/Z$ 
  can be confined, but also 
  sets a criterion for the maximum energy 
  a nucleon can acquire  
  through stochastic acceleration in a region 
  involving magnetic confinement in the acceleration process 
  (cf. the Hillas criterion~\cite{Hillas1984ARAA}).   
  A direct scaling of the expression 
  in Eq.~\ref{eq:zeta_nu} 
  gives a corresponding expression 
   for a relativistic electron/positron of energy $E_{\rm e}$: 
\begin{align} 
\zeta_{\rm L,e} \approx 8.5 \times 10^{-11}\ 
   \left(\frac{E_{\rm e}}{1\;\!{\rm GeV}} \right)  
  \left(\frac{\mathcal D}{1\;\!{\rm Mpc}} \right)^{-1}  
  \left(\frac{ {\mathcal B}}{10\;\!{\rm nG}} \right)^{-1} \ .  
\label{eq:zeta_e}
\end{align}    
Protons with $\zeta_{\rm L,p}\;\! 
  (= \zeta_{\rm L,nu}\vert_{(A/Z)=1}) > 1$ 
  and electrons/positrons with $\zeta_{\rm L,e} > 1$  
  are able to break magnetic confinement.  
In most astrophysical situations, 
  the energies of electrons 
  would not greatly exceed ${\rm PeV}$ levels. 
Thus, $\zeta_{\rm L,e} \ll 1$ would be expected 
  in filament environments.

%
\subsection{Magnetic-field configurations}  
\label{subsec:field_configuration} 

While there is a consensus that cosmic filaments are magnetised, 
  little is known about 
  the properties of their magnetic fields, including their strength, 
     their global and local topology     
     and their connectedness 
     to the internal magnetic fields of 
 lower-hierarchical systems 
     linked to the filaments   
     (e.g. groups or clusters) 
     or embedded within them (e.g. field galaxies). 
Direct measurements of magnetic fields 
  beyond galaxy cluster scales  
  is a great technical challenge. 
Currently, 
 only loose constraints can be derived for
 the strengths of the magnetic fields 
  in filaments or in voids, 
  and we practically have no reliable information 
  about the field topology or how magnetic fields in filaments and voids 
  interface.  
The strengths of magnetic fields in voids  
  are inferred to be below nG, 
  based on the directional anisotropy 
   observed in ultra-high-energy (UHE) cosmic rays 
  \citep[][]{Hackstein2016MNRAS}, the absence of a trend 
  in the rotation measure (RM) of distant radio sources 
  over redshift \citep[][]{Pshirkov2016PhRvL}, and the lack of any clear detection of pair echos or halos from distant $\gamma$-ray point sources due to deflected  electromagnetic cascades~\citep[e.g.][]{Ackermann2018ApJS, Vovk2023PhRvD}. 
 Magnetic fields in filaments are estimated to 
 be around 30~nG 
 \cite[][]{Vernstrom2021MNRAS,Carretti2023MNRAS}. While we may derive such constraints on their strengths based on arguments 
  invoking thermodynamics 
  (e.g. energy equipartition between particles and the magnetic field) 
  or radiative processes (such as synchrotron emission 
  and/or Compton scattering), 
  we still do not have a reliable means 
    to determine the configurations 
    of magnetic fields in filaments 
    theoretically or observationally. 
    
The confinement and transport 
  of energetic particles in filaments 
  is sensitively dependent on 
  the topological properties and effective strengths of the magnetic fields. 
Modelling how energetic charged particles behave 
  in cosmological filament environments 
  is not a trivial generalisation of 
  modelling how energetic charged particles behave 
  in diffuse media on stellar and galactic scales. 
This is partly due to the sheer scale of cosmological filaments, 
  but also because the fate of energetic particles 
  is dependent on properties of filaments 
  at different stages of their cosmological evolution. 
It is also partly due to 
  the connectedness of the filament magnetic field to the magnetic fields of the embedded galaxies (where energetic charged particles originate) and 
   the clusters/superclusters 
  where the filaments terminate. 
Filaments can therefore be considered  
  as a part 
  of a larger ecological web, 
  in which particles are energised, 
  destroyed, converted and recycled.

%
%
\begin{figure}[H] 
\vspace*{0.2cm}
\begin{adjustwidth}{-\extralength}{0cm}
\centering
\includegraphics[width=5.75cm]{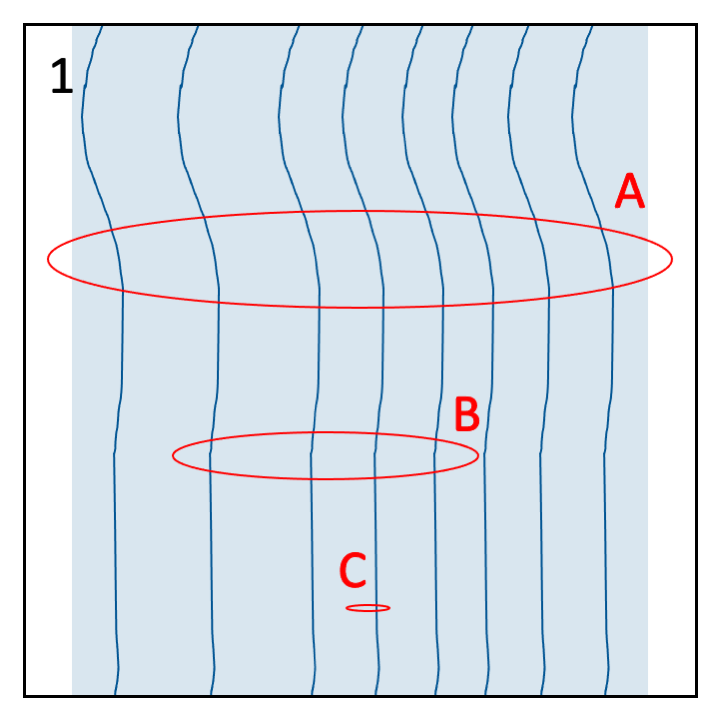} 
\includegraphics[width=5.75cm]{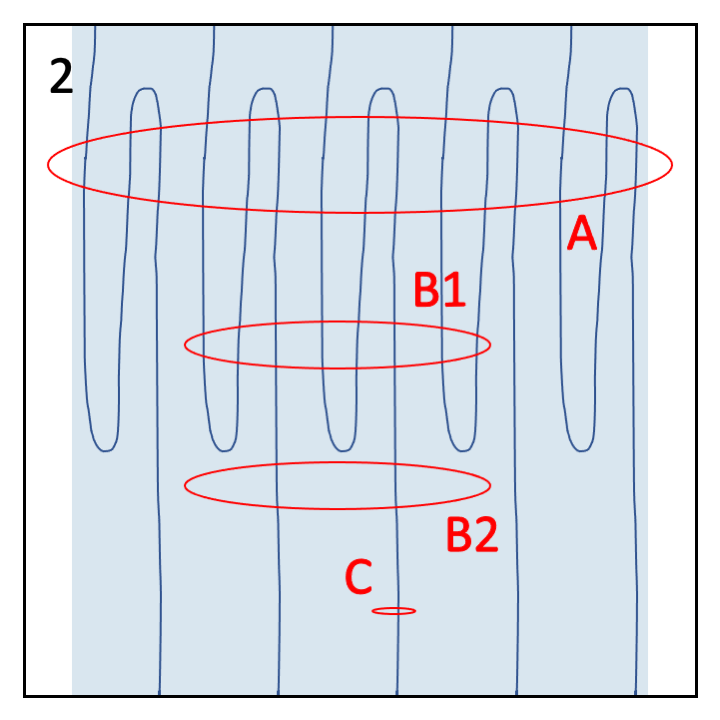} 
\includegraphics[width=5.75cm]{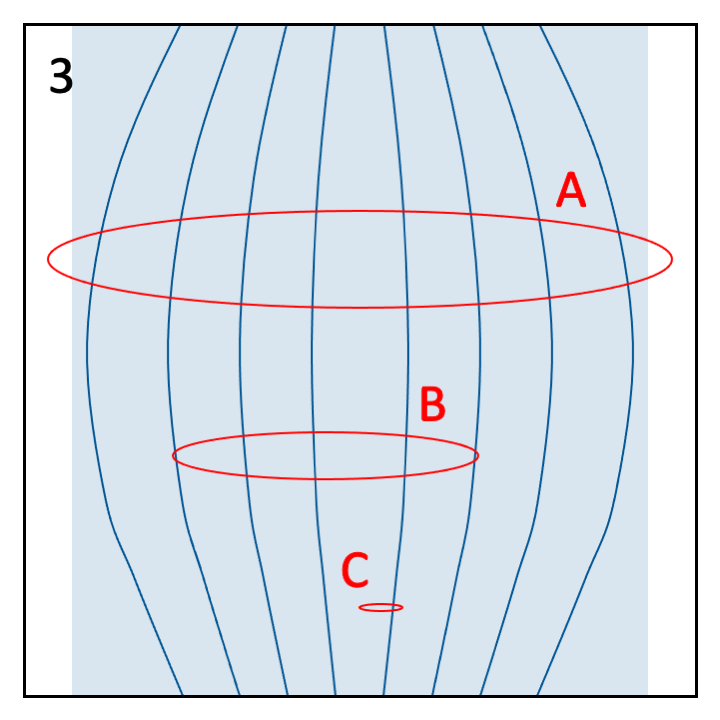} \\ 
\includegraphics[width=5.75cm]{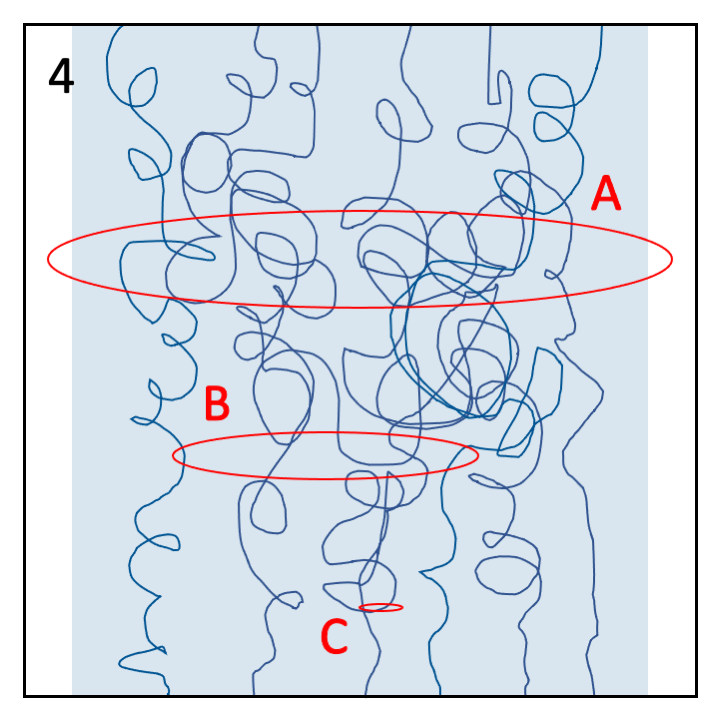} 
\includegraphics[width=5.75cm]{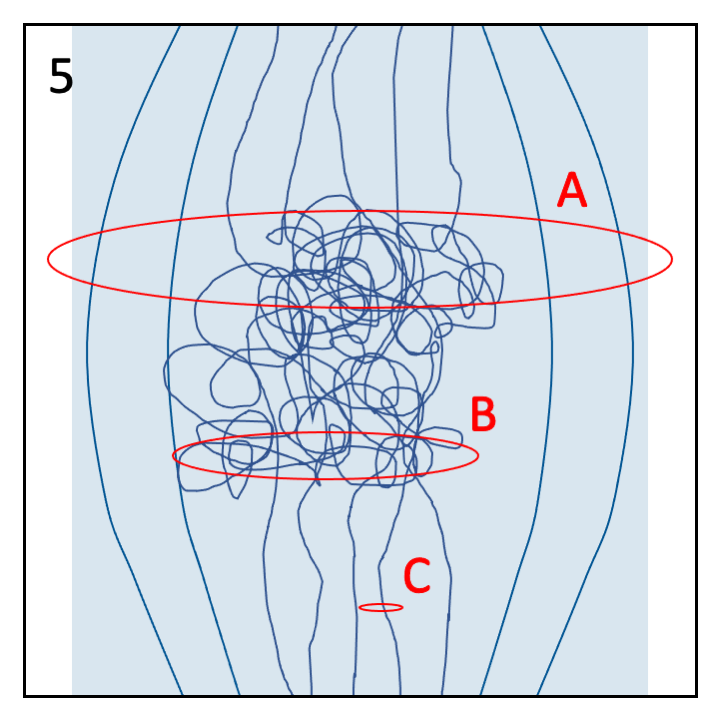} 
\includegraphics[width=5.75cm]{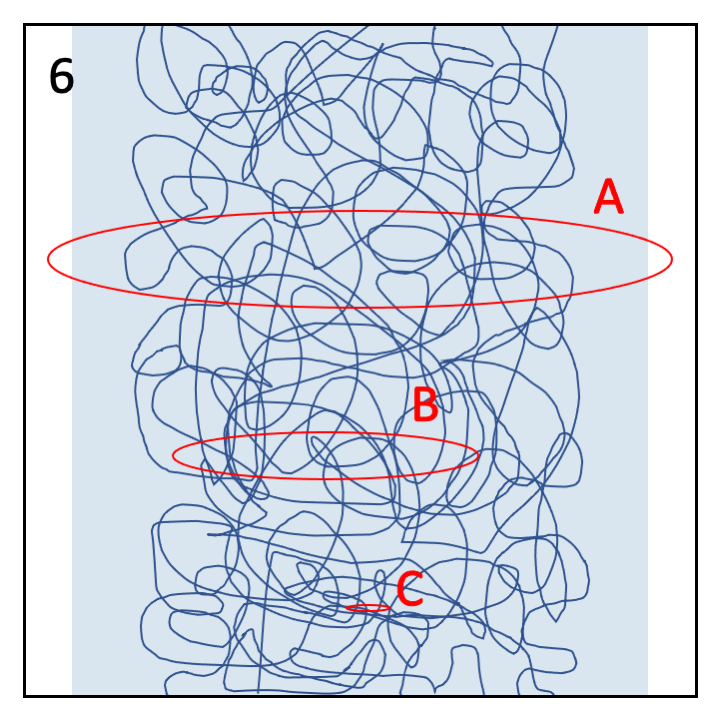} 
\end{adjustwidth}
\caption{Schematic illustrations of 
  various generic situations   
  for the confinement and transport 
  of energetic charged particles 
  in segments of cosmological filaments 
  (denoted by the shaded light blue region 
  in each panel) 
  threaded by magnetic fields 
  (represented by curved dark blue lines).  
The gyration of the charged particles 
  in the magnetic fields 
  may be classified into 
  three regimes 
  represented by the red ellipses (gyration orbits), 
  not to scale,  
  marked with A, B (or B1 and B2) 
  and C. 
The regimes  
   are defined according to 
   the value of $\zeta_{\rm L, p}$, 
   by setting in Equation~\ref{eq:zeta_nu}  
   (for $A/Z = 1)$ 
   the domain extent ${\mathcal D}$ 
   as the width of the filament 
   and characteristic magnetic field strength ${\mathcal B}$ 
   as that given by Equation~\ref{eq:cal_B}. 
The orbit types labelled as A correspond to 
 cases where $\zeta_{\rm L, p} > 1$,   
  the orbit types labelled as B (and B1/B2) correspond to 
 cases where $\zeta_{\rm L, p} \lesssim 1$ 
  and the orbit types labelled as C 
  correspond to cases 
  where $\zeta_{\rm L,p}\ll 1$.   
The magnetic fields in the panels 
  have two components, 
  a component 
    with a relatively well-ordered 
    large-scale 
    field configuration within the region 
    (i.e. the presence of a guided field) 
  and a component 
    with disordered smaller-scale field configuration. 
Panel 1 shows a filament segment where 
  the magnetic field is dominated 
  by an ordered large-scale field 
  without orientation reversal. 
Panel 2 shows a filament segment 
  where the magnetic field is dominated 
  by an ordered large-scale field 
  but there are field reversals warping 
  parts of the segment.  
Panel 3 shows a filament segment 
  where the dominant component of  
  the magnetic fields 
  have a bottle configuration. 
Panel 4 shows a filament segment 
  where the disordered small-scale field 
   components are as prominent as the  
   the large-scale well ordered component, 
   which has an orientation parallel to the filament. 
Panel 5 shows a filament segment 
  where a prominent disordered small-scale 
  field component is embedded 
  inside a large-scale component 
  with a bottle configuration.  
Panel 6 shows a filament segment 
  where the well-ordered large-scale 
  field component is absent, 
  leaving only the presence of 
  a disordered small-scale field component. 
Note that type B orbits 
  are split into two subgroups, B1 and B2, in panel 2, 
  with the former in the folded field region 
  and the latter outside the folded field region.  
} 
\label{fig:B_fig01}
\end{figure}
%
%

We now lay out a conceptual framework, 
  which provides a basis 
  for meaningful discussion 
  of the fate of energetic particles 
  in their complex interplay  
  with filament environments.  
Magnetic fields undoubtedly regulate 
  the destruction, conversion and recycling of energetic particles  
  in cosmological filaments.     
Self-gravitating astrophysical systems  
  generally have round shapes if they are gravitationally-supported 
  (e.g. stars, elliptical galaxies and galaxy clusters)
  or disk shapes if they are supported by angular momentum 
  (e.g. proto-stellar systems and disk galaxies).   
Filaments are the only self-gravitating systems 
  that can persist over long timescales 
  (cf. Hubble time), 
  with elongated structures. 
The unique geometry of filaments 
  allows them to harbour  
  a coherent magnetic structure along their symmetry axis 
  over length-scales substantially 
  larger than their thickness.

With this in mind, we may 
  consider a filament segment 
  of characteristic linear size $\ell$ 
  within a filament of length $\mathcal L$, 
  where ${\mathcal L} \gg \ell$. 
This segment is 
  part of a long filament section, which   
  is permeated by a magnetic field. 
The magnetic field can be decomposed 
  into a relatively well ordered 
  large-scale component 
  ${\boldsymbol B}_{\rm g}$ 
  and a disordered smaller-scale component 
  ${\boldsymbol B}_{\rm s}$, i.e. 
\begin{align} 
  {\boldsymbol B} 
  = {\boldsymbol B}_{\rm g} + {\boldsymbol B}_{\rm s} \ .    
\end{align}   
The large-scale component 
  serves as the field substratum. 
It can be further decomposed 
 into two orthogonal components:  
 ${\boldsymbol B}_{\rm g} 
   = {\boldsymbol B}_\parallel 
  + {\boldsymbol B}_\phi$,    
  where ${\boldsymbol B}_\phi$
  is the global toroidal component, 
   presumably supported by a large-scale 
   current flow along the filament, 
  and ${\boldsymbol B}_\parallel$ 
  is the parallel component 
  contributing to the global directional  
  magnetic flux 
  aligned with the orientation of the filament. 
In the absence of a large-scale current, 
  ${\boldsymbol B}_{\rm g} ={\boldsymbol B}_\parallel$. 
Without losing generality, 
  we simply assume that 
   the large-scale current is negligible, 
   i.e.  
   $|{\boldsymbol B}_\parallel| \gg |{\boldsymbol B}_\phi|$. 
The smaller-scale disordered field component,  
  ${\boldsymbol B}_{\rm s}$, 
  contributes to the rest of 
  the total magnetic field  
  in the filament segment. 
We assume that the small-scale field component 
  is statistically isotropic 
  (i.e. it has no preferential orientation),   
  and statistically homogeneous  
  (i.e. it is invariant under spatial translation).  
As the magnetic field is divergence free, 
  the magnetic fluxes 
  of the small-scale disordered field component  
  passing through an arbitrary surface constructed within the filament,  
  with an area 
  similar to or 
  larger than the cross section of a galaxy,    
  is statistically zero. 
These properties of an arbitrary surface 
  do not hold 
  for the large-scale field component, 
  though its magnetic flux 
  entering and leaving 
  a volume element within the filament segment 
  is strictly zero. 
We shall demonstrate in later sections 
  that such differences between 
  the large-scale and the small scale field components 
  would have subtle effects  
  on the properties and fate 
  of baryonic cosmic rays in the filaments.  

Figure~\ref{fig:B_fig01} shows six examples of generic 
  field configurations in a filament segment. 
These configurations are reasonable representations  
  in regions without sizeable substructures 
  such as galaxies, groups or clusters. 
They are constructed based on the relative strengths 
 of ${\boldsymbol B}_{\rm g}$ and ${\boldsymbol B}_{\rm s}$, 
 with additional considerations 
 such as large-scale fields along the filament 
 linking a pair of galaxy clusters close by. 

In observational and simulation studies, 
  the properties of magnetic fields 
  are often quantified 
  with a coherence length $\lambda_{\rm B}$ 
  \citep[see e.g.][]{Durrer2013A&ARv,Han2017ARAA}. 
For the convention of magnetic-field decomposition 
  that we have adopted, 
  ${\boldsymbol B}_{\rm g}$ 
  would have a larger value for $\lambda_{\rm B}$ 
  while ${\boldsymbol B}_{\rm s}$ 
  would have a smaller value. 
Note that ${\boldsymbol B}_{\rm g}$ 
  in the field configuration shown in panels 1, 2 and 3 all have 
  $\lambda_{\rm B}\approx{\mathcal D}$, 
  yet the variations in their structures 
  would lead to very different effects on the fate of charged particles. 
Thus, in a more thorough formulation, 
  a comprehensive description beyond 
   coherence length parameters 
  $\lambda_{\rm B}$ would be necessary, though 
  our approach is still important 
  to provide useful insights 
  into the properties of magnetic fields 
  in interstellar medium (ISM)  
  or IGM. 
For clarity in the field topological analysis, 
  hereafter 
  we do not consider 
  the coherence length parameter explicitly unless otherwise stated.  
  We instead describe 
  the magnetic fields directly 
  using a two-component  
  ${\boldsymbol B}_{\rm g}$-${\boldsymbol B}_{\rm s}$ 
  decomposition.

In the first three cases shown in Fig.~\ref{fig:B_fig01}  
  the small-scale disordered field component 
  is negligible compared 
  with the large-scale field component.  
When $|{\boldsymbol B}_{\rm g}| 
  \gg |{\boldsymbol B}_{\rm s}|$,   
  the magnetic field is relatively ordered  
  and aligned with the filament orientation (see Panel 1). 
If there is a strong flow along the filaments 
  (e.g. in channelled accretion), 
  Rayleigh Taylor instabilities could develop, 
  which may lead to a field folding 
  in some regions in the filament segment 
  (see Panel 2). 
A magnetic bottle field configuration could develop in the region  
  between two embedded galaxies in a filament 
  (see Panel 3). 
This requires that the flow within the filament 
  is dominated by bulk motion instead of turbulent motion. 
This can occur naturally, and can be understood as follows. 
The turbulent speeds $v_{\rm turb}$ 
  in IGM 
  are in the range $10-50\;\!{\rm km~s}^{-1}$, 
  as inferred from observations 
  \citep[e.g.][]{Xu2020ApJ,Bolton2022MNRAS}.    
It is generally considered that  
  filaments contain substantial amount of WHIM, 
  which has a temperature of $\sim 10^5 - 10^7\;\!{\rm K}$. 
If we take a gas temperature  
  $T\sim 10^6\;\!{\rm K}$, 
  it gives a sound speed 
  $c_{\rm s} \sim 70\;\!{\rm km~s}^{-1}$.  
Thus, the IGM turbulence is generally subsonic. 
The flows in filaments are not pressure supported, 
  and it has been argued 
  that shocks are present in filaments 
  \citep[][]{Vernstrom2023SciA}. 
This implies that
  the bulk flow speed $v_{\rm bulk}$ 
  along the filament is supersonic, 
  i.e. $v_{\rm bulk} > c_{\rm s} > v_{\rm turb}$ 
  (at least in certain regions within a filament). 
  
In the last three cases shown in Fig.~\ref{fig:B_fig01}, 
   the small-scale disordered field component 
  is non-negligible compared 
  with the large-scale field component, 
  at least in some regions, 
  or even dominates over the large-scale field component. 
When the two field components are comparable, 
  the magnetic field in the filament segment 
  would share the characteristics of the two components 
  (see Panel 4). 
While the field lines appear to be entangled, 
  they do not mask the global orientation of the 
  directed large-scale field component. 
There are also situations 
  where disordered magnetic fields 
  are generated inside 
  a magnetic-bottle field configuration (see Panel 5)  
  by turbulence motion of the ionised gas (plasma) 
  or energetic particles trapped inside,  
  or by the presence of a shock 
  caused by colliding outflows from two galaxies.  
If the filament segment does not have strong bulk flow, 
  it is possible that the disordered field component 
  dominates (see Panel 6), 
  reflecting the turbulent nature of the gas in the region.

%
\subsection{Particle confinement}  
\label{subsec:particle_trapping} 

To determine $\zeta_{\rm L,nu}$ and $\zeta_{\rm L,e}$ 
  for nucleons or electrons of given energies, 
  we need to assign values 
  for ${\mathcal D}$  and ${\mathcal B}$ 
   in Equations~\ref{eq:zeta_nu} and \ref{eq:zeta_e}.  
For the value of ${\mathcal D}$, 
 the filament thickness can be adopted as    
  inferred from observations, such as stacking.  
The value for ${\mathcal B}$ 
  is currently not measurable 
  directly from observations. 
Instead, it can be derived 
  under certain assumptions,   
  such as energy equipartition 
  between the magnetic field and non-thermal particles, 
  if the energy content  
  of the emitting particles 
  can be determined from observations. 
Different field configurations 
  could give the same ${\mathcal B}$. 
Fig.~\ref{fig:B_fig01} 
  show examples of magnetic field configurations, 
  with different combinations 
  of ${\boldsymbol B}_{\rm g}$ and 
  ${\boldsymbol B}_{\rm s}$,  
  in a two-component representation. 
For uncorrelated ${\boldsymbol B}_{\rm g}$ 
  and ${\boldsymbol B}_{\rm s}$, 
  the characteristic field strength is then 
\begin{align} 
  {\mathcal B} & = 
    \sqrt{ \langle {{\boldsymbol B}_{\rm g}}^*\cdot 
      {\boldsymbol B}_{\rm g}  \rangle 
   +  \langle {{\boldsymbol B}_{\rm s}}^*\cdot 
   {\boldsymbol B}_{\rm s}  \rangle 
   + 2 \langle {{\boldsymbol B}_{\rm g}}^*\cdot 
    {\boldsymbol B}_{\rm s}  \rangle } \nonumber \\ 
  & \approx \sqrt{ |{\boldsymbol B}_{\rm g}|^2 
   + |{\boldsymbol B}_{\rm s}|^2 }  \ . 
\label{eq:cal_B}
\end{align} 

For protons in a filament segment 
  with $\{ {\mathcal D},{\mathcal B} \}$,   
  $\zeta_{\rm p}\;\! 
  (= \zeta_{\rm L, nu}\vert_{(A/Z = 1)})$  
  will depend only on $E_{\rm p}$, the proton energy. 
If we set ${\mathcal B} \sim 30\;\!{\rm nG}$   
  (a value similar to the field estimated for large-scale filament 
  \citep[see][]{Carretti2022MNRAS}), 
  protons with energies of $3.5\times 10^{19}\;\!{\rm eV}$,  
  $1.1 \times 10^{18}\;\!{\rm eV}$ 
  and $3.5\times 10^{16}\;\!{\rm eV}$, 
  would have gyration radii of 
  1~Mpc, 40~kpc and 1~kpc,  
  respectively.  
The protons may therefore be sorted into three groups, 
  (i) $\zeta_{\rm L,p} > 1$, 
  (ii) $\zeta_{\rm L, p} \lesssim 1$ 
  i.e. slightly smaller than but of the same order as 1,
  and (iii) $\zeta_{\rm L,p} \ll 1$, 
  according to their energies. 
  The protons in these groups 
  have their gyration radii 
  larger than, comparable with but slightly smaller, 
  and much smaller than the thickness of the associated filament, 
  respectively.  

We now illustrate how differently  
  the protons in the three groups would behave 
  for different magnetic field configurations 
  (even though they may share the same value for ${\mathcal B}$).  
We ignore the drift of the protons  
  along the filament for the time being 
  and focus on their motion  
  perpendicular to the filament orientation.  
Schematic illustrations of 
  the gyration orbits, 
  labelled as A, B (B1/B2) and C 
  for the proton groups (i), (ii) and (iii) respectively,  
  are shown in Fig.~\ref{fig:B_fig01}. 

As gyration orbit type A 
  exceeds the filament thickness, 
  the protons with this orbit
  will not be confined to the filament 
  in all cases shown in Fig.~\ref{fig:B_fig01}. 
Gyration orbit type B is slightly 
  smaller than the filament thickness, 
  and proton confinement 
  would occur in this case, at least in principle. 
However, non-uniformity 
  in the magnetic field  
  will cause these protons 
  to drift across filament. 
Protons of gyration orbit type B 
  can therefore only be confined and survive 
  in the filament 
  over a timescale that depends 
  on the competition 
  between particle interactions  
  and particle escape through cross-field particle diffusion \cite[see, e.g.][]{Xu2013ApJ}. 
Protons of gyration orbit type C 
  will be trapped in the filaments. 

The filament segment shown in Panel 1 
  is the generic situation. 
The lack of ${\boldsymbol B}_{\rm s}$ in this case 
  implies weaker diffusion 
  for protons with gyration orbit type B, 
  compared to the other field configurations 
  in Fig.~\ref{fig:B_fig01}. 
The well-ordered ${\boldsymbol B}_{\rm g}$ 
 aligned along the filament segment also implies 
 that protons with gyration orbit type C 
  could be channelled out 
  from the segment through drift along 
  the field direction. 

The filament segment in Panel 2 
  has a field folding region.  
The magnetic field reversal  
   alters the confinement ability of the region,  
  and hence that of the filament segment.   
Protons that should otherwise have been retained   
 according to their energies $E_{\rm p}$ and the characteristic field strength 
  ${\mathcal B}$ in the filament segment
  (such as those with gyration orbit type B2),  
  can now break free of confinement. 
Field folding therefore  
   opens up a back-door 
   for protons (those with gyration orbit type B1) 
   to leave a filament 
   through direct escape 
   or by fast tracking cross-field diffusion. 
  
Magnetic mirroring in Panel 3 
 will restrict the drift of the protons  
  with gyration orbits 
  smaller than the thickness of 
  the filament segment  
  and will retain them in the region. 
The effect is stronger for protons 
  with gyration orbit type C than 
  those with gyration orbit type B.   
Protons of gyration orbit type B 
  could drift out of the filament 
  through cross-field diffusion, 
  similar to the situation shown in Panel 1. 
Protons leaving the segment 
  in the direction along the filament 
  would be suppressed 
  by magnetic mirroring, 
  unlike those in Panel 1.  

The field configurations 
  in Panels 4 and 6 
  can be considered 
  as continuations of that shown in Panel 1, 
  as the strength of ${\boldsymbol B}_{\rm s}$ 
  relative to that of ${\boldsymbol B}_{\rm g}$ 
  increases. 
This increase 
  will enhance the diffusion 
  of protons with gyration orbit type B 
  out of the filament segment. 
This in turn reduces the confinement time   
  of these protons. 
Protons with gyration orbit type C 
  are not affected greatly, 
  unless the corresponding $r_{\rm C}$, 
   the radius of gyration orbit type C, 
   strongly violates the condition 
\begin{align} 
  \int_{2\pi/{\ell}}^{2\pi/r_{\rm C}}{\rm d}^3{\boldsymbol k} \ 
    \left|\tilde{{\mathcal B}}({\boldsymbol k}) \right|^2   
  \gg  \int_{2\pi/r_{\rm C}}^\infty 
    {\rm d}^3{\boldsymbol k} \ 
    \left|{\tilde{\mathcal B}}({\boldsymbol k}) \right|^2   \ , 
\end{align}
  where 
  $\tilde{{\boldsymbol B}}({\boldsymbol k})$ 
  is given by the Fourier transform 
\begin{align} 
  \tilde{{\mathcal B}}({k})
   \approx \frac{1}{2} \ \int_{-1}^{1} {\rm d}\mu \int_0^{\ell/2} 
     {\rm d} r \ r^2\  {\mathcal B}(\boldsymbol r) \ 
   e^{-{\rm i}\;\!{\mu k\;\! r}}  \  
\end{align} 
for isotropic ${\boldsymbol B}_{\rm s}$. 

The magnetic field configuration in Panel 5 
  has the same large-scale field as that in Panel 3.   
The presence of the small-scale disordered field component 
  ${\boldsymbol B}_{\rm s}$ 
  in the magnetic bottle  
  set by the large-scale field component ${\boldsymbol B}_{\rm g}$  
  will not affect 
  the global confinement of protons 
  with gyration orbit type C.  
These protons will be reflected back  
  by magnetic mirroring,  
  once they diffuse out from the 
  region with ${\boldsymbol B}_{\rm s}$. 
The presence of the small-scale disordered field component 
  ${\boldsymbol B}_{\rm s}$ 
   facilitates the diffusion of 
  protons with gyration orbit type B 
  across the filament 
  where confinement is regulated only 
  by ${\boldsymbol B}_{\rm g}$.

\section{Filament ecology}  
\label{sec:filament_ecology}

%
\subsection{Filament interfaces} 
\label{subsec:interface} 

Filaments are not isolated structures.   
  They are an integral part of an ecological system  
  for the production, transportation, conversion and destruction 
  of energetic particles.  
The interaction between the filaments and other components 
  in this ecosystem determines  
  the composition, spectrum, 
  and fate of energetic particles. 
The interfaces of these components are gateways 
  for energetic particles.  
We may broadly divide these interfaces 
  into three basic classes:  
   filament-void interfaces, 
  filament-cluster/supercluster interfaces 
  and filament-galaxy interfaces.    
This classification is not an artificial construct, 
  and we will illustrate that these interfaces 
  act as sieve for energetic particles entering and exiting  filaments. 

\begin{figure}[t]
\vspace*{0.0cm}
\begin{adjustwidth}{-\extralength}{0cm}
\centering
\includegraphics[width=10cm]{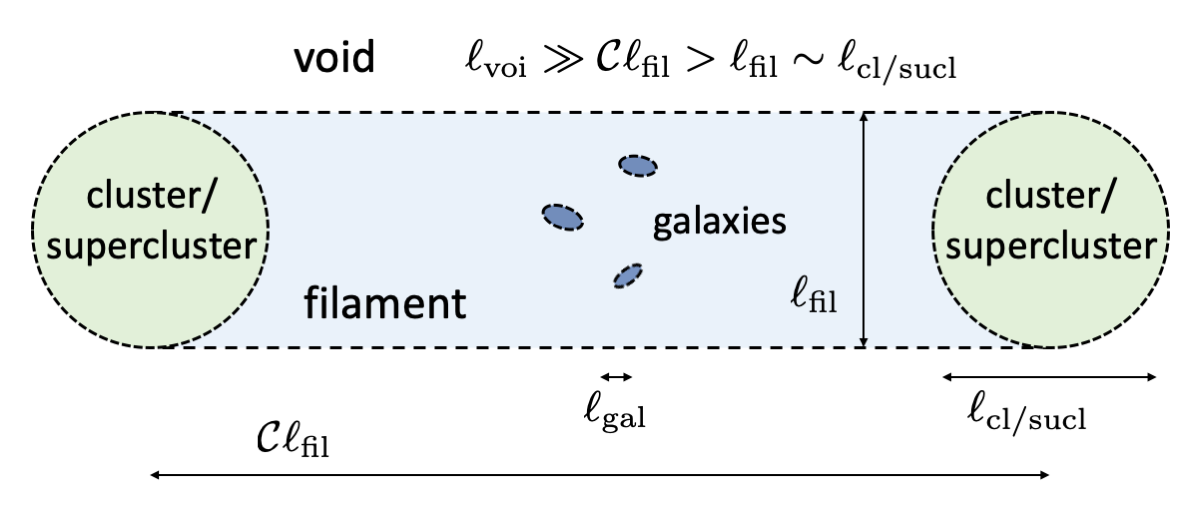} 
\end{adjustwidth}
\caption{A schematic illustration to show 
  the interfaces of a filament with a void, 
   two clusters/superclusters, 
   and several galaxies embedded in it. 
The size bars indicating the length-scales 
  $\ell_x$ \ 
  ($\;\! x\in \{{\rm fil},\;\!{\rm voi},\;\!{\rm cl/sucl},\;\!{\rm gal}\;\!\}$) 
  are not exactly to scale. 
The filament has an elongated shape, hence 
  the filament aspect-ratio scaling variable ${\cal C} >1$.}  
\label{fig:topology}
\end{figure}

From a geometrical perspective,  
  interfaces broadly correspond 
  to two magnetic field topologies between three classes of objects of different sizes  
  relative to the size of the filament segments involved: (i) 
  the linear sizes of voids $\ell_{\rm voi} \gg \ell_{\rm fil}$,   
  (ii)  the linear sizes of clusters and superclusters\footnote{Galaxy clusters 
    are not all virialized objects 
    \citep[see e.g.][]{Xu2000ApJ}
    and superclusters are not virialised. 
  Superclusters cannot be described 
    by simple geometrical shapes, 
    such as spheres or ellipses. 
   For example, the Laniakea Supercluster 
    \citep[][]{Tully2014Natur}
    does not have a well defined shape. 
    It is elongated, and presumably threaded 
      by many filaments at different locations. 
   The term `linear size' 
   of the cluster and supercluster here 
     is therefore used in a loose context. }
  $\ell_{\rm cl/sucl} \approx 
  \eta\;\! \ell_{\rm fil}$ 
  (where $\eta \sim [{\mathcal O}(1)]$),  
  and (iii) the linear sizes of galaxies  $\ell_{\rm gal} \ll \ell_{\rm fil}$. 
Up to this point, 
  we have been using the linear size of a filament segment, $\ell$,  
  and the thickness of the filament, $\ell_{\rm fil}$, interchangeably  
  (see Sec.\ref{subsec:field_configuration}) 
  in the context that 
  $\ell \longleftrightarrow \ell_{\rm fil} < {\cal C}\ell_{\rm fil}$,      
  where ${\cal C}>1$ is the aspect-ratio scaling variable 
  of the filament.    
These three interface classes 
  are defined 
  by the topological nature 
  of how filaments are associated with the other components 
  in the ecosystem. 
The properties of interfaces 
  do not follow simple scaling relations 
  according to the sequence of galaxies, 
  cluster/superclusters and voids. 
Moreover,  
  we cannot simply scale their physical properties, 
  such as the magnetic field strength 
  and the thermal content of gas 
according to the characteristic sizes of the component structures.      
As illustrated in Fig.~\ref{fig:topology}, 
  galaxies are enclosed by filaments, and filaments 
  and clusters/superclusters are enclosed by voids, 
  but clusters/superclusters are linked by filaments.  

%
%
%
%
  
\begin{table}[h] 
\vspace*{0.25cm}
\caption{Passages of cosmic rays through interfaces between filament eco-components.}
\begin{tabular}{llcccc}
\hline
Interface type        & \hspace*{0.0cm} &  A           & B          & C          &  \\
\hline
 Void to filament &   & \tikzxmark  & \checkmark & \checkmark &  \\
Filament to void  &   & \checkmark  & ?          & \tikzxmark &  \\
\hline
Cluster/supercluster to filament&  & \tikzxmark  & ?          & ?          &  \\
Filament to cluster/supercluster&  & \tikzxmark  & \checkmark & ?          & \\
 \hline 
Galaxy to filament&   & \tikzxmark  & \checkmark & ?         &  \\
Filament to galaxy&   & \tikzxmark  & ??         & ?         & \\
\hline
\end{tabular} 
\vspace*{0.15cm} \\ 
\noindent{\footnotesize{\textbf{Notes}: 
 A table to summarise 
   the prospects of 
   particles with given gyration orbits 
 to transfer 
   between filaments and voids, 
   filaments and clusters/superclusters,  
   and filaments and embedded galaxies. 
A, B and C correspond 
  to the gyration orbit types A, B and C 
  in Fig.~\ref{fig:B_fig01}. 
`\checkmark' and `\tikzxmark' 
  denote which cases could and could not transfer particles via   
 the described pathway, 
  respectively. 
`?' denotes that transfer through the described  
  pathway is subject to the efficiency of 
  diffusion across the magnetic field,  
  in competition with other relevant processes  
  e.g. the survival of particles in the presence  
  of pp or p$\gamma$ interactions. 
`??' denotes that there could be complications in the transfer of particles through the described pathway 
  caused by other factors, 
  such as the presence of a magnetic barrier 
  in the filament-cluster/supercluster interface,   
  and/or the diffusion 
  of particles through the magnetic field internal 
  to the systems.} }  
\label{tab:gateway}
\end{table}

This classification is based on 
  geometrical and topological considerations. 
  The magnetic fields in filaments  
 introduce an additional layer of complexity. 
If we ignore the magnetic field configurations  
  associated with the interfaces for the time being 
  and adopt the gyration orbits of particles 
  (as described in Sec.~\ref{subsec:particle_trapping})  
  as a reference, 
  it can immediately be 
  seen how energetic particles are sieved differently 
  by the three interfaces 
  (which is summarised in Table~\ref{tab:gateway}). 
It is clear that  
  voids can accommodate 
  the most energetic particles 
  (with gyration orbit type A) 
  without difficulty. 
It is also apparent that 
  filament-void interfaces are a one-way opening 
  for low-energy particles 
  (with gyration orbit type C).  
These particles can freely enter a filament from a void,  
  but they are unable to break the confinement from a filament 
  to escape to a void. 
Generally, filaments can accept the particles 
  of gyration orbit type B 
  from galaxies, clusters/superclusters or voids.  
Whether these structures can retain particles  
  depends on how far the particles are able to diffuse across the filament 
  before undergoing a hadronic interaction 
  (see Sec.~\ref{sec:hadronic_interactions}).

%
%
\begin{figure}[t]
\vspace*{-0.cm}
\begin{adjustwidth}{-\extralength}{0cm}
\centering
\includegraphics[width=5.75cm]{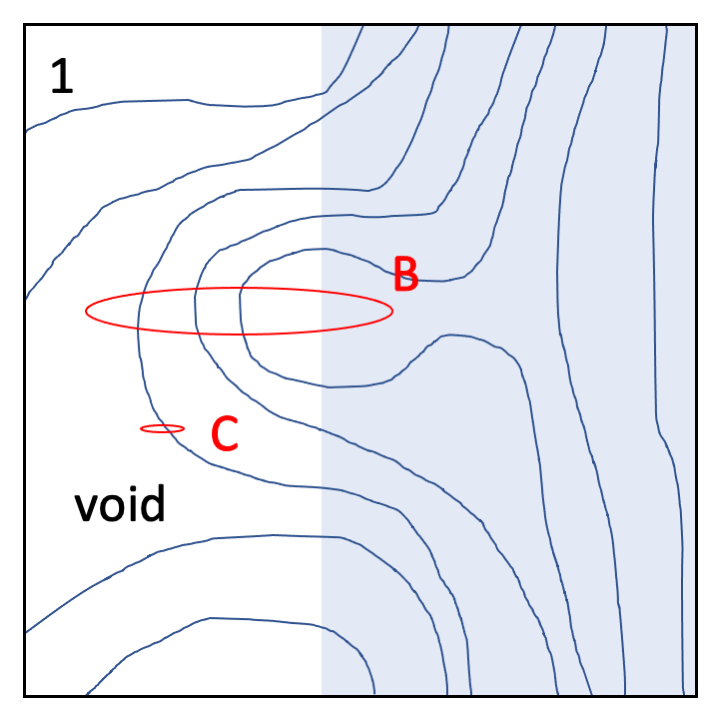} 
\includegraphics[width=5.75cm]{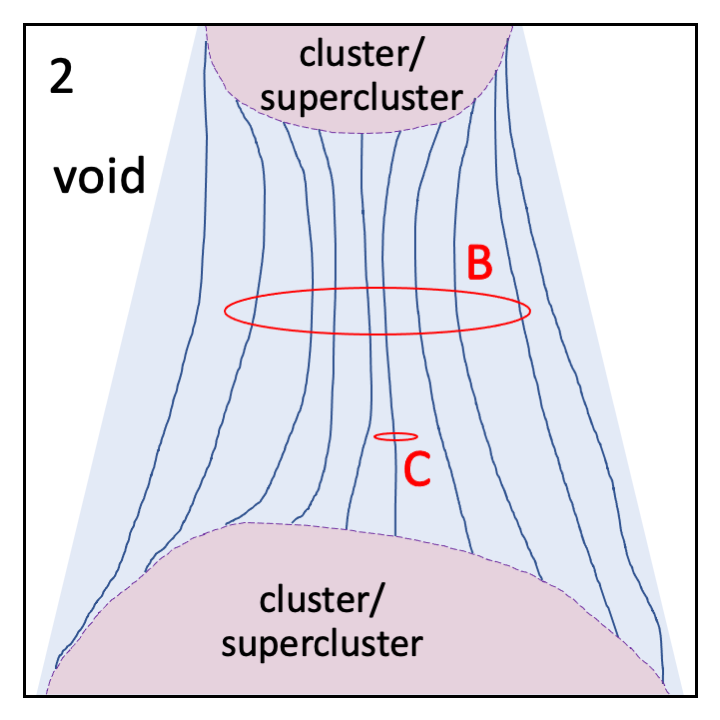} 
\includegraphics[width=5.75cm]{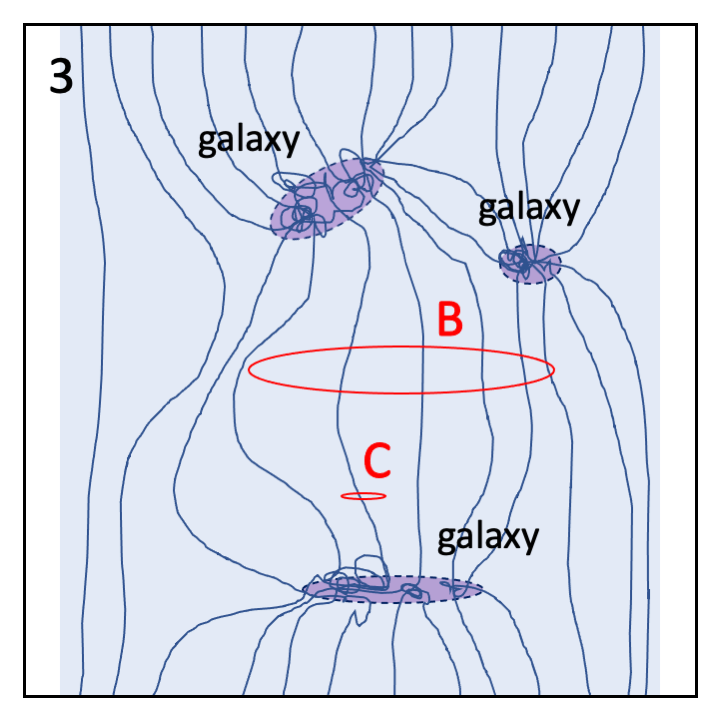} 
\end{adjustwidth}
\caption{Schematic illustrations 
  of three specific situations 
  for the confinement and propagation 
  of energetic particles 
  in filament environments.    
Panel 1 shows the closed and open magnetic field 
  lines in the interfacing regions 
  of a filament and a void. 
Panel 2 shows the filament connecting 
  two clusters/superclusters, 
  where the filament magnetic field lines  
  have a concave configuration 
(note that we distinguish the filament here 
 from the inter-cluster bridges 
 between two merging galaxy clusters).   
Panel 3 shows the interweaving structure 
  of magnetic field lines connected 
  between a group of galaxies embedded in a filament. 
The ellipses B and C are defined  as those in Figure~\ref{fig:B_fig01}
  (with respect to a characteristic magnetic field 
  and the size of the filament segment considered).  
    } 
\label{fig:B_fig02}
\end{figure}
%
%

Magnetic fields are divergence free, 
 so the open field lines of a filament magnetic field 
 must thread through the interfaces with other eco-components. 
Fig.~\ref{fig:B_fig02} illustrates  
  schematically 
  examples of the topologies of global, 
  relatively ordered magnetic fields 
  ${\boldsymbol B}_{\rm g}$
  associated with the interfaces 
  between a filament segment and 
  a void (Panel 1), two clusters/superclusters (Panel 2) 
  and several embedded galaxies (Panel 3).  

The two kinds of magnetic field topologies 
  associated with filament-void interfaces  
  are open and closed field lines.  
This is analogous to the magnetic field configuration of stars 
  \citep[see e.g.][]{Keppens2000ApJ,Wiegelmann2017SSRv}.  
In stellar magneto-spheres, 
 closed field lines bound dead zones 
   where charged particles become trapped,  
  while open field lines thread the wind zones 
   where charged particles are allowed to escape.  
In the filament-void interface 
 (Panel 1, Fig.~\ref{fig:B_fig02}), 
  particles of gyration orbit types B and C 
  leak out from the filament and 
  will be deflected by the closed magnetic field lines
   back to the filament 
   (cf. the dead zone in a stellar magneto-sphere). 
If the particles leak out from  
  regions threaded by open magnetic field lines,  
  they will continue to drift outwards without confinement  
  (cf. the wind zone in a stellar magneto-sphere).   
Particles in voids 
  with gyration orbit types B and C 
  can be channelled into filaments 
  when they are magnetically captured, 
  regardless of whether the field line is open or closed. 

At the interfaces between 
  a filament and its linked clusters/superclusters,  
  the ${\boldsymbol B}_{{\rm g}\parallel}$ component 
  should be non-negligible  
  as the open field lines
  in the filament will terminate there 
  (see illustration in Panel 2, Fig.~\ref{fig:B_fig02}). 
Energetic particles could therefore drift across the interface 
  along the ${\boldsymbol B}_{{\rm g}\parallel}$ component 
  from a filament into a cluster/supercluster 
  (or vice versa).  
The magnetic fields in filaments are of $\sim 10\;{\rm nG}$ 
 \citep{Vazza2021Galax,Carretti2023MNRAS}. 
The magnetic fields in the core of clusters 
  have strengths of up to $\sim 10\;\mu{\rm G}$ 
    \citep{Govoni2004IJMPD,Vacca2018Galax}, 
    but are weaker at their outskirts. 
The gyration orbit of a particle will therefore shrink 
  by almost three orders of magnitude 
  when it enters a cluster from a filament. 
Particles with gyration orbit type B in a filament 
  can easily be captured by a cluster/supercluster at the interface. 
By contrast, 
  particles that diffuse from a cluster/supercluster 
  into a filament to end up as particles 
  with a gyration orbit type B 
  would have had much smaller gyration orbits  
  when they were inside the cluster/supercluster. 

The magnetic fields at a filament-cluster/supercluster interface 
  may have some interesting properties 
  based on their topological analysis. 
The magnetic field at the interface must have 
  a strong toroidal field component perpendicular 
  to ${\boldsymbol B}_{{\rm g}\parallel}$ 
  and/or a strong smaller-scale disordered field component,  
  even if the filament segment 
  linking to the cluster/supercluster 
  lacks a small-scale disordered field component.  
This can be understood as follows. 
The linear extent of the interface would be 
  $\ell_{\rm cl/sucl}$, 
  which is of the same order as $\ell_{\rm fil}$. 
Over this length-scale, the strength of the magnetic field 
  would need to increase from  
  $10~{\rm nG}$ levels to $\sim 10~\mu{\rm G}$. 
The magnetic field energy 
  cannot increase substantially  
  by squeezing or bending  
  ${\boldsymbol B}_{{\rm g}\parallel}$ in the filament 
  as $\ell_{\rm fil} \sim \ell_{\rm cl/sucl}$. 
Thus, the options to accommodate 
  this disparity between the field strengths 
  in the filament and the linked cluster/supercluster 
  (if adopting the observed values of filaments and clusters available 
  at the moment) 
  are (i) the presence of a strong toroidal field component  
  (not resulting from deformation of the
  ${\boldsymbol B}_{{\rm g}\parallel}$ component), 
  (ii) a strong localised small-scale disordered field component,  
  or (iii) both of these. 
Whether or not particles  
  would need to overcome this magnetic barrier when entering or exiting 
  the cluster/supercluster would depend on their  
 diffusion and scattering 
   when 
  crossing this barrier. This 
  is not considered  
  in our qualitative analysis 
  using gyration orbits, 
  and needs additional consideration in future, more comprehensive 
  modelling studies. 

Different to filament-cluster/supercluster interfaces, 
  the linear extent of the filament-galaxy interfaces 
  is determined by the sizes of the galaxies.  
  It is independent of the thickness of the filament segment containing the galaxies 
  (see Panel 3, Fig.\ref{fig:B_fig02}). 
As the sizes of galaxies are significantly 
  smaller than the thickness of their host filament, 
  i.e. $\ell_{\rm gal} \ll \ell_{\rm fil}$, 
  and galactic magnetic fields 
  \citep[several tens of $\mu{\rm G}$, see][]{Beck2015A&ARv}
  are at least three orders of magnitude stronger than 
  the magnetic fields of filaments \cite[tens of ${\rm nG}$][]{Carretti2023MNRAS}, 
  particles with gyration orbit type B in a filament  
    would be captured by galaxies  
    through a ballistic collision  
    instead of a diffusion or a diffusive drift process 
    in the filament magnetic field
    (regardless of the galactic field configuration).  
The capture probability of these particle may be estimated as 
  $P_{\rm capture} \sim ({\Upsilon}/{\mathcal C}) 
  ({\ell_{\rm gal}}/{\ell_{\rm fil}})^2$, 
  where the structural factor $\Upsilon\  (\sim {\cal O}(1))$ 
  depends on the aspect ratio of the galaxy, and  
  the relative orientation and location of the galaxy in its host filament.  
For $\ell_{\rm gal} \lesssim 30\;\!{\rm kpc}$ and  
  $\ell_{\rm fil} \sim 2~{\rm Mpc}$, as ${{\mathcal C} > 1}$, 
  the capture probability of 
  these filament particles by a galaxy would be well below 1\%.  
Particles with gyration orbit type C in a filament would 
  enter a galaxy through diffusive drift. Their  
   orbit would shrink very significantly after entering the galaxy. 
Particles with gyration orbit type C originating from 
  an embedded galaxy should have much smaller gyration orbits. 

The question now to be asked is  
  whether filament-galaxy interfaces 
  would have similar magnetic barriers to those expected 
  for filament-cluster/supercluster interfaces, 
  when the filament magnetic fields have a significant 
  non-toroidal large-scale component. 
The non-toroidal filament magnetic field, 
  if not bypassing the galaxy, 
  would terminate at the filament-galaxy interface. 
At the interface, the field lines of these components 
  could connect with local poloidal open field lines 
  or toroidal open field lines from within 
  the galaxy. 
As an illustration we may consider  
 a disk galaxy of radius 
  $15~{\rm kpc}$ embedded within  
 a filament of thickness $2~{\rm Mpc}$. 
Suppose there is a magnetic field line bundle  
  with a radius of $0.5~{\rm Mpc}$ 
  threaded into an embedded galaxy.  
The magnetic field strength 
  in the arms of nearby spiral galaxies 
  is $\sim 10~\mu{\rm G}$ 
  \citep[see e.g.][]{Beck2015A&ARv}. 
It follows that 
\begin{align} 
  \frac{\Phi_{\rm fil}}{\Phi_{\rm gal}} 
    \sim \left( \frac{10~{\rm nG}}{10~\mu{\rm G}}  \right)  
      \left( \frac{0.5~{\rm Mpc}}{15~{\rm kpc}} \right)^2 
      \sim 1  \ , 
\end{align}
 where $\Phi_{\rm fil}$ is the magnetic flux 
 of the filament magnetic field bundle that threads onto the disk galaxy, 
 and $\Phi_{\rm gal}$ 
   is the magnetic flux of the open-line magnetic field  
     from the galaxy. 
This implies that, unlike the filament-cluster/supercluster interfaces,   
 filament-galaxy interfaces 
  can easily accommodate filament magnetic fields,  
  even if ${\boldsymbol B}_{{\rm g}\parallel}$ dominates. 
If a galaxy has a strong central starburst, 
  the magnetic field in the   
  starburst nucleus could exceed $\sim 100~\mu{\rm G}$ \citep[see, e.g.][]{Beck2015A&ARv, Peretti2019MNRAS}. 
Then, the large-scale ordered field of the host filament cannot connect smoothly 
  to the galactic magnetic field. 
  This implies the presence of a toroidal field component and/or 
  a small-scale disordered field component.  
In this case, a magnetic barrier for the charged particles 
  would be formed, similar to the situation at  
 the filament-cluster/supercluster interface.  
 We note that this filament interface would also be 
  where the boundary of the CGM is located. 
If the CGM gas 
  is formed by ejecta 
   from the starburst core 
   of the galaxy, 
  the magnetic fields threaded through it 
  could have a substantial small-scale disordered field component, or perhaps even 
  a toroidal field component.  

%
\subsection{Journey, life-cycle and fate 
 of energetic particles}  

The question behind most current studies 
  of the transport of cosmic rays 
  and their interactions   
  is how the medium 
  would affect their transport processes 
  and properties.  
An ecosystem, however, 
  consists of complex webs 
  of multiply-connected components.  
The question posed above 
  is therefore not particularly meaningful, 
  given the multiplicities  
  of energetic particle interactions 
  and the interconnection between 
  the components of the ecosystem that 
   energetic particles would encounter.  
Posing a grander question 
  of how the web of components associated with cosmic filaments 
  would affect the global transport processes of cosmic rays 
  and the final and intermediate properties 
  of the particles within the components of the system 
  will, however, 
  make the problem impossible to tackle. 
An alternative approach 
  is to put focus on the individual particles 
  and investigate 
  how they react to the individual components in the system they would encounter.  
We therefore apply the information obtained 
  by studying the journey of individual particles 
  to deduce the life-cycle and fate 
  of energetic cosmic rays  
  in filaments 
  and to derive useful insights that can then be applied to the broader astrophysical context. 

We start with this question: 
  what will happen to a particle, say a proton, 
  of energy $E_{\rm p}$ 
  starting its journey from a location within a galaxy, a cluster, a supercluster or a filament at a particular cosmological epoch?  
Without losing generality, 
  we consider three protons at energies of 
  $10^{12}\;\!{\rm eV}$, 
  $10^{16}\;\!{\rm eV}$ and $10^{20}\;\!{\rm eV}$. 
The threshold energy for  
  pion production  
  p$\gamma$ processes 
  in astrophysical environments  
  is above $10^{16}{\rm eV}$ 
  (see Sec.~\ref{sec:hadronic_interactions}).  
It is considered possible to accelerate 
  particles to energies 
  as high as $10^{20}~{\rm eV}$ 
  in astrophysical systems, 
  without violating the Hillas criterion 
  \citep[][]{Hillas1984ARAA}.
The three energies we have chosen
  bracket the energy ranges for 
  two regimes: the first is the 
  transport of cosmic rays that do not undergo significant hadronic p$\gamma$ processes 
  ($10^{12}-10^{16}\;\!{\rm eV}$), while the second is the transport of cosmic rays that have a possibility 
  to undergo a hadronic p$\gamma$ process 
  ($10^{16}-10^{20}\;\!{\rm eV}$).  
We consider four initial locations 
  for the protons: 
  the filament itself, a very large supercluster, a cluster, 
  and a disk galaxy 
  (which may or may not have starburst).   
The assigned characteristic 
  magnetic field strengths 
  are $10~{\rm nG}$ for the filament, 
  $1~\mu{\rm G}$ for superclusters and clusters,  
  and $10~\mu{\rm G}$ and $100~\mu{\rm G}$ 
  respectively for the disk galaxies with and without starbursts. 
For completeness, 
  we also assign a value of $10^{-15}\;\!{\rm G}\ (=10^{-6}\;\ 1{\rm nG})$ 
  \citep[][]{Beck2013MNRAS,Samui2018MNRAS}  
  for the magnetic fields in voids.   

%
\subsubsection{Particles starting 
  from a filament}  
\label{subsubsec:start_filament}  

%
%
\begin{figure}[t]
\begin{adjustwidth}{-\extralength}{0cm}
\centering
\includegraphics[width=15cm]{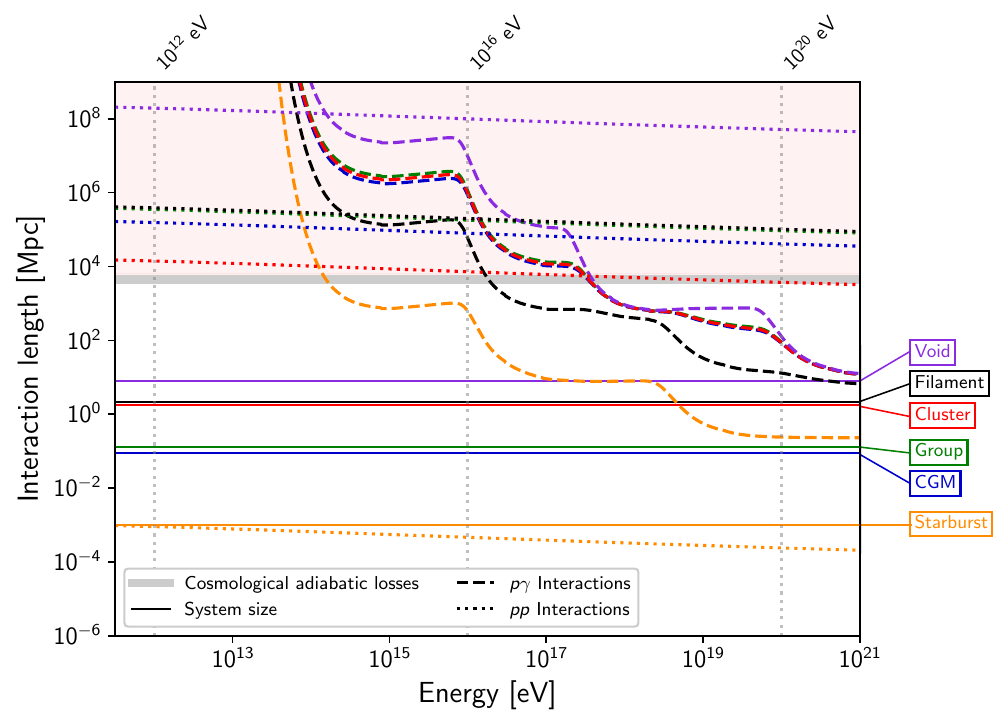} 
\end{adjustwidth}
\caption{Interaction lengths of protons 
  undergoing pp and p$\gamma$ interactions 
  in cosmic voids (shown by purple lines), 
filaments (black lines) and their internal structures: galaxy clusters (shown in red), groups of galaxies (green), the CGM, of a galaxy (blue), and a starburst galaxy (orange). All path lengths are calculated at the current epoch ($z=0$). The length-scale for proton adiabatic losses due to cosmological expansion is shown for comparison. 
Interaction lengths above this scale  
  (indicated by the pink shaded region) 
  are not of astrophysical consequence, 
   but are shown for completeness. 
The parameters adopted for these calculations are shown in Appendix~\ref{app:AAAAA}, where filament conditions are taken as their central values, while the characteristic size is taken as the filament outskirts, i.e. $\sim$ 2 Mpc. Vertical lines mark the three proton energies discussed in the main text. }
\label{fig:interactions_components}
\end{figure}    
%
%

Consider first that the three protons 
  are produced in a filament 
  at the present epoch ($z=0$).  
With the assigned filament magnetic field 
  of $10\;\!{\rm nG}$, 
  their Larmor radii $r_{\rm L}$ 
  are $3.33\times 10^{17}\;\!{\rm cm}$ ($\sim 0.1~{\rm pc}$)   
  (for $10^{12}\;\!{\rm eV}$), 
  $3.33\times 10^{21}\;\!{\rm cm}\ 
    (\sim 1~{\rm kpc})$  
  (for $10^{16}\;\!{\rm eV}$)  
  and $3.33\times 10^{25}\;\!{\rm cm}\ 
    (\sim 10~{\rm Mpc})$  
  (for $10^{20}\;\!{\rm eV}$). 
The thickness of a filament 
  at the present epoch 
  would be in the range 
  $(1-4)~{\rm Mpc}$, 
  which is less than 
  $3\times 10^{25}\;\!{\rm cm}$. 
Thus, only the proton of $10^{20}{\rm eV}$ 
  would be able to escape 
  from the filament to the void. 
  The other two 
  would be confined within the filament. 
  
If the $10^{20}\;\!{\rm eV}$ 
  proton manages to leave 
  the filament and escape into the void, 
  its Larmor radius 
  would be $> 10^{32}~{\rm cm}~(\sim 10^{4}~{\rm Gpc})$, 
  if we take the magnetic field 
  in the void to be 
  $10^{-15}\;\!{\rm G}~
  (=10^{-6}\;\!{\rm nG})$. 
It practically free-streams 
  away and would not return to 
  the filament it originated from. 
At the current epoch,  
  it could travel over a distance 
  of a few tens Mpc   
  as collisions 
  with CMB photons 
  degrade its energy through photo-pair and 
 pion-production 
  (see Sec.~\ref{sec:hadronic_interactions}). 
If its energy drops to 
  around $10^{16}\;\!{\rm eV}$, 
  it would still undergo
  free-streaming propagation 
  as its Larmor radius 
  would be $\sim 10^{28}\;\!{\rm cm}~(\approx3.2~{\rm Gpc})\ 
  \gg {\cal C}\ell_{\rm fil} > \ell_{\rm fil}$. 
It cannot be captured magnetically 
  by a filament 
  unless it hits it directly, or unless  
  it collides with a  
  gravitational substructure 
  associated with an embedded component of the filament. 
At the energies 
  $\sim 10^{16}{\rm eV}$, 
  the proton cannot undergo 
  p$\gamma$ processes to produce 
  pions, nor can it lose energy 
  efficiently through any other process. 
The proton is practically frozen at 
  this energy, wandering in the void. 
If its energy  
  drops to about $10^{\rm 12}\;\!{\rm eV}$  
  during a p$\gamma$ interaction 
  (though this is unlikely, as  
  almost all its energy will be passed 
  to pions in a single collision with 
  a CMB photon), 
  it could then be magnetically captured 
  by a filament, 
  as it would then have a Larmor radius of 
  about $3\times 10^{24}\;\!{\rm cm}~(1~{\rm Mpc})$, 
  i.e. is comparable to $\ell_{\rm fil}$. 
  
Filament-like structure 
  might have begun to appear 
  at redshifts as high as $z\sim 4$, 
  and filaments continue 
  to evolve to become the current form. 
Energetic protons would respond differently 
  when encountering a filament at $z=2$ 
  and a modern day filament, 
  because of filament evolution 
  and the cosmological conditions. 
A $10^{20}\;\!{\rm eV}$ proton 
  would have an interaction length 
  of about 10~Mpc at $z =2$ 
  and a fit more 
  at $z=0$
  (Fig.\ref{fig:hadronic_ints}) 
  to p$\gamma$ interactions, and would again lose energy through this process 
  when colliding with a CMB photon. 
If the proton's energy drops by a factor 
  of 10 through photo-pair and pion-production, 
  it can be confined 
  even by a thin filament 
  of thickness of $1~{\rm Mpc}$. 
When the proton is deflected 
  back to a filament, 
  it will collide with a 
  photon in the CMB 
  or a local radiation field 
  again 
  until its energy drops 
 below the p$\gamma$ interaction 
  energy threshold.  
  
Lower-energy protons 
 confined in filaments 
 would scatter or diffuse 
 within them, 
 depending on the coherence length-scale 
 of the ordered magnetic field. 
For the case 
  of a cosmic ray proton captured by a cluster, 
  if there is a magnetic barrier 
  as described in Sec.~\ref{subsec:interface} 
  at the filament-cluster interface, 
  it will take addition time 
  to diffuse into the cluster. 
When protons enter a cluster/supercluster 
 or an embedded galaxy, 
 their Larmor radii will expand roughly according to the scaling 
 $r_{\rm L,x} = r_{\rm L,fil}
 [(\langle |{\boldsymbol B}|^2 \rangle_{\rm x})^{1/2}
 /{\mathcal B}]$. 
Here, ${\rm x}\in
  \{ {\rm gal,\;\! cl/sucl} \}$ 
  and ${\mathcal B} = (\langle |{\boldsymbol B}|^2 
  \rangle_{\rm fil})^{1/2}$.
As the energies of these protons  
  are below the photo-pair and pion-production energy thresholds, they will not participate in 
 p$\gamma$ processes. However  
  they may lose a small fraction of their 
  energy through the direct production 
  of electron/positron pairs 
  when colliding with CMB photons 
  in the filament. If they are captured by 
  a galaxy (see Fig.~\ref{fig:interactions_components}), they may 
 also participate in pp interactions. 
  

%
\subsubsection{Particles starting from a cluster  
  or a supercluster}  
\label{subsubsec:start_cluster} 

For a magnetic field of $1\;\!\mu{\rm G}$, 
 the Larmor radii of 
  protons with energies of 
  $10^{12}\;\!{\rm eV}$, 
  $10^{16}\;\!{\rm eV}$ 
  and $10^{20}\;\!{\rm eV}$ 
  are 
  $3.3\times 10^{15}\;\!{\rm cm}$ 
   $3.3\times 10^{19}\;\!{\rm cm}$ 
   ($\sim 10~{\rm pc}$), 
    $3.3\times 10^{23}\;\!{\rm cm}$ 
    ($\sim 0.1~{\rm Mpc}$),  
    respectively. 
The size of a cluster is a few Mpc, 
  and the linear extent of a supercluster 
  can exceed 100~Mpc.~\footnote{The 
  Laniakea supercluster and the 
  Saraswati supercluster are among 
  the biggest known superclusters to date. 
  They are estimated to have 
  a (maximum) linear extent of 
   $\sim 160\;\!{\rm Mpc}$  
  \citep{Tully2014Natur} 
   and $> 200\;\!{\rm Mpc}$  
  \citep{Bagchi2017ApJ}, respectively.}

The sizes of clusters and superclusters 
  are significantly larger than 
  the Larmor radius 
  of the $10^{20}\;\!{\rm eV}$ proton 
  in $\mu{\rm G}$ level 
  magnetic-field strengths.   
As magnetic-field components 
  in clusters or superclusters  
  are expected to have   
  coherence lengths that are smaller than the sizes of clusters/superclusters, 
  the proton with an energy of 
  $10^{20}\;\!{\rm eV}$ 
  may diffuse out from its host cluster/supercluster  
  into a filament or a void  
  and avoid being captured 
  by a galaxy. 
This could occur in the current epoch ($z=0$), 
  as the p$\gamma$ 
  interaction length 
  for a $10^{20}\;\!{\rm eV}$ proton 
  is a few tens of Mpc (see Fig.~\ref{fig:interactions_components}). 
The interaction length, however drops 
  with an increase in redshift, 
  and at $z=2$ 
  the corresponding p$\gamma$ interaction length 
  becomes smaller than 1~Mpc, 
  implying that photo-pair and 
 pion-production arising in collisions 
  with CMB photons can substantially 
  degrade the energy of the proton. 
The proton may still leak out from a cluster 
  with a size of a few Mpc, 
  but it is unlikely to be able to   
   escape intact from a supercluster 
  with size of 100~Mpc.   

The two protons with lower energies of $10^{12}$ and $10^{16}\;\!{\rm eV}$  would be    
  trapped within their cluster/supercluster 
  of origin. 
Their energies are below the p$\gamma$ interaction  
  energy threshold (see Fig.~\ref{fig:interactions_components}), 
  so they practically become fossilised at this energy with 
 pair-production being unable 
  be to cool them rapidly. 
However, 
  the Larmor radii of these protons 
  is sufficiently small 
  that they could become entangled 
  and advected by 
  cluster-scale flows (e.g. in mergers),  
  or in strong AGN outflows 
  or AGN induced large-scale bubbles. 
Otherwise, they will diffuse 
  around until being captured by 
  a galaxy within the cluster/supercluster. 

%
\subsubsection{Particles starting from a disk galaxy}  
\label{subsubsec:start_cluster}

Observations have shown that 
  the magnetic fields of disk galaxies 
  often have a relatively well ordered pattern 
  \citep[e.g.][]{Han2006ApJ,Beck2015A&ARv,Han2017ARAA}. 
In a magnetic field of $10\;\!\mu{\rm G}$, 
  protons with energies of 
  $10^{12}\;\!{\rm eV}$, 
  $10^{16}\;\!{\rm eV}$ 
  and $10^{20}\;\!{\rm eV}$ 
  have Larmor radii of 
  $3.3\times 10^{14}\;\!{\rm cm}$, 
  $3.3\times 10^{18}\;\!{\rm cm}\ (\sim 1~{\rm pc})$, 
  $3.3\times 10^{22}\;\!{\rm cm}\ (\sim 10~{\rm kpc})$, 
  respectively. 
The diameter of a Milky-Way like galaxy 
  is about $30~{\rm kpc}$ 
  and the scale-height of the galactic disk 
  would be of about $(1-2)\;\!{\rm kpc}$  
  \citep[see e.g.][]{Rix2013A&ARv,Hayden2017A&A}. 

A $10^{20}\;\!{\rm eV}$ proton, 
  if they are produced 
  in violent environments, 
  such as from a gamma-ray burst 
  \citep[see e.g.][]{Vietri1995ApJ}, 
  an AGN \citep[see e.g.][]{PAO2008APh}
  or even a weakly accreting black hole, 
  could easily stream out of the galactic disk,   
  unless it first collides with an ISM baryon 
  or a photon from the radiation field 
  of some bright stellar objects.  
The p$\gamma$ interaction length 
  of $10^{20}\;\!{\rm eV}$ protons 
  is significantly larger than 
  the diameter of the galaxy, 
 hence 
  there is little chance it  
  can collide with a CMB photon 
  before leaving its galaxy of origin. 
The final destination of this fugitive proton, 
  which retains its energy, 
  may be in a cluster, a supercluster 
  or a filament, 
  and its fate will be similar to 
  the respective $10^{20}\;\!{\rm eV}$ protons 
  as described 
  in Sec.~\ref{subsubsec:start_filament} 
  and~\ref{subsubsec:start_cluster}. 
The lower-energy protons 
  will be confined 
  by the disk magnetic field 
  and end up undergoing a pp interaction 
  when colliding with a baryon,  
  or a p$\gamma$ interaction 
  when colliding with a photon 
  from the stellar radiation field. 

The situation would be different  
  if the disk galaxy 
  has a strong outflow 
  driven by the starburst 
  \citep[see e.g.][]{Anchordoqui2020PhRvD} 
  from within. 
First, these galaxies may have 
  a stronger magnetic field 
  \cite[see][]{Beck2015A&ARv}. 
Second, disk galaxies  
  with strong outflows 
  would have a different topology   
  to disk galaxies 
  without an outflow,  
  especially in the presence 
  of large-scale open field lines 
  which extend into the galactic halo 
  (see e.g. NGC 4631, NGC 891 and M 82~\citep[][]{Mora2013A&A,Krause2009RMxAC,Pattle2021MNRAS}). 
For a magnetic field of $100\;\!\mu{\rm G}$, 
 the Larmor radii of 
  protons with energies of 
  $10^{12}\;\!{\rm eV}$, 
  $10^{16}\;\!{\rm eV}$ 
  and $10^{20}\;\!{\rm eV}$ 
  are 
  $3.3\times 10^{13}\;\!{\rm cm}$, 
  $3.3\times 10^{17}\;\!{\rm cm}\;\! 
    (\sim 0.1\;\!{\rm pc})$, 
  $3.3\times 10^{21}\;\!{\rm cm}\;\!
  (\sim 1\;\!{\rm kpc})$, respectively.  
A proton with an energy of $10^{20}\;\!{\rm eV}$ 
 would be scattered by the galactic magnetic field 
 and leave the galaxy,  
 provided it survives  
 pion producing hadronic collisions 
 with ambient baryons or photons.   
The fate of this proton is similar 
 to its corresponding fugitive proton 
 from the disk galaxy without an outflow.  
The proton with an energy of 
 of $10^{16}\;\!{\rm eV}$ 
 and of $10^{12}\;\!{\rm eV}$  
 could be advected out of the galaxy, 
 practically intact, 
 if it is entangled 
 in the magnetic field carried by the outflow.  
Otherwise, 
  the proton of such a low energy  
  will reside within its galaxy of origin 
  until it collides with a baryon 
  and lose its energy 
  in the particle production cascades 
  (see Fig.~\ref{fig:interactions_components}).

\section{Astrophysical implications}  
\label{sec:implications}

%
\subsection{Filaments as cosmic ray highways 
  and fly papers}  
\label{subsec:highway} 

The ability of filaments to retain 
  energetic particles  
  gives them a very special role 
  in the transfer of cosmic rays  
  on a cosmological scale. 
In most current studies of cosmological-scale 
  cosmic ray transport,  
  all material outside galaxies 
  is broadly referred to as the ``IGM''.  
This is treated as 
  a single static agent, 
  with which cosmic rays interact 
  as they propagate through the Universe. 
This ``IGM" is generally described 
  in terms of certain variables in a statistical manner, 
  which are either inferred from observations 
  or extrapolated from simulations.  

In the previous section, 
we demonstrated that 
  energetic protons starting from galaxies, 
  clusters, superclusters and filaments 
   each have a different life journey,   
  marked by their identity. 
  Their fate depends on 
  where they come from, 
  where they were born, 
  and how much energy they initially acquired.  
With a qualitative, heuristic analysis 
  of the journey and fate of energetic protons 
  originating from 
  a filament, 
  a cluster and a supercluster linked to a filament, 
  and a galaxy embedded inside a filament, 
  we have been able to 
 derive insights 
  into various aspects of cosmic ray ecology within  
 filament environments. 
Filaments are special large-scale structures  
  which have direct contact with  
  all key eco-components - 
  voids, cluster/superclusters 
  and galaxies. 
This connectedness  
  implies that filaments 
  play an important role 
  in mediating and regulating  
   cosmological cosmic ray transfer 
  in a manner that has much more physical complexity 
  than a simple diffusion/scattering scenario 
  can easily accommodate. 

Our analysis has shown that 
  filaments are cosmological-scale highways which confine cosmic rays and channel them between 
clusters, superclusters and galaxies.  
An energetic particle escaping  
  from a cluster is not expected to 
  have a good chance to ballistically hit   
  a neighbouring cluster 
  when it is free streaming 
  in vast inter-cluster space  
  where the magnetic fields are weak 
  (below $\sim 10^{-15}\;\!{\rm G}$ 
  in cosmic voids). 
However, if the energetic particle enters a filament, 
  it will be magnetically channelled 
  towards another cluster at the 
  other end of the linking filament. 
The situation is the same 
  for cosmic ray particles leaving a galaxy or a galaxy group 
   embedded within a filament.  
These are channelled towards nodes as their destinations\footnote{Apart from clusters 
  and superclusters, 
  galaxy groups can also serve as nodes of large scale filaments. 
  In a recent observation, the M 101 galaxy group 
   was identified as a node in a nearby cosmic filament 
   \citep{Karachentseva2023A&A}.}.  

Voids enclose filaments, 
  and filaments cannot serve as highways 
  to channel cosmic rays 
  across void-filament interfaces. 
Cosmic rays breaking filament confinement 
  would have a Larmor radius larger than  
  the thickness of a filament. As such, they 
  would be scattered when they 
  encounter another filament along their 
  path~\cite[e.g.][]{Kotera2008PhRvD, Owen2023PhRvD}. 
These cosmic rays need to undergo 
  a `transformation' 
  before they can be captured 
  and retained by a filament.   
Cosmic ray protons with energies above 
  the p$\gamma$ interaction threshold 
  can degrade to become lower-energy protons 
  through their collisions with CMB photons. 
When their energies drop to  
  $\sim 10^{18}\;\!{\rm eV}$, 
  their Larmor radii $r_{\rm L}$ will be  
  $\sim 3\times 10^{23}{\rm cm}\ 
  (\approx 100\;\!{\rm kpc})$ 
  for a magnetic field of 
  $|{\boldsymbol B}| \approx 10\;\!{\rm nG}$.  
A $10^{18}\;\!{\rm eV}$ proton 
  can then be magnetically captured 
  when intercepted by a filament. 
This means that filaments 
  can act as cosmic ray `fly paper', 
  targeting lower-energy particles. 
Given that 
  the volume filling factor of filaments in the Universe  
  is estimated to be about 10\% 
  \citep[see e.g.][]{Tempel2014MNRAS}, the chance of a streaming cosmic ray to collide with a filament is not 
   negligible. Once a cosmic ray has been 
   captured by a filament, 
   it can then undergo p$\gamma$ interactions inside, degrading its energy to a few $10^{16}\;\!{\rm eV}$.   
 These captured cosmic rays   
  will eventually become frozen at some energy   
  $\lesssim 10^{16}\;\!{\rm eV}$ 
  over the timescale  
   it takes them to traverse the filament.  
   
%
\subsection{Cumulative calorimetry} 
\label{subsec:calorimetry} 

\begin{figure}[H] 
\begin{adjustwidth}{-\extralength}{0cm} 
\centering 
\vspace*{-0.25cm}
\includegraphics[width=13.5cm]{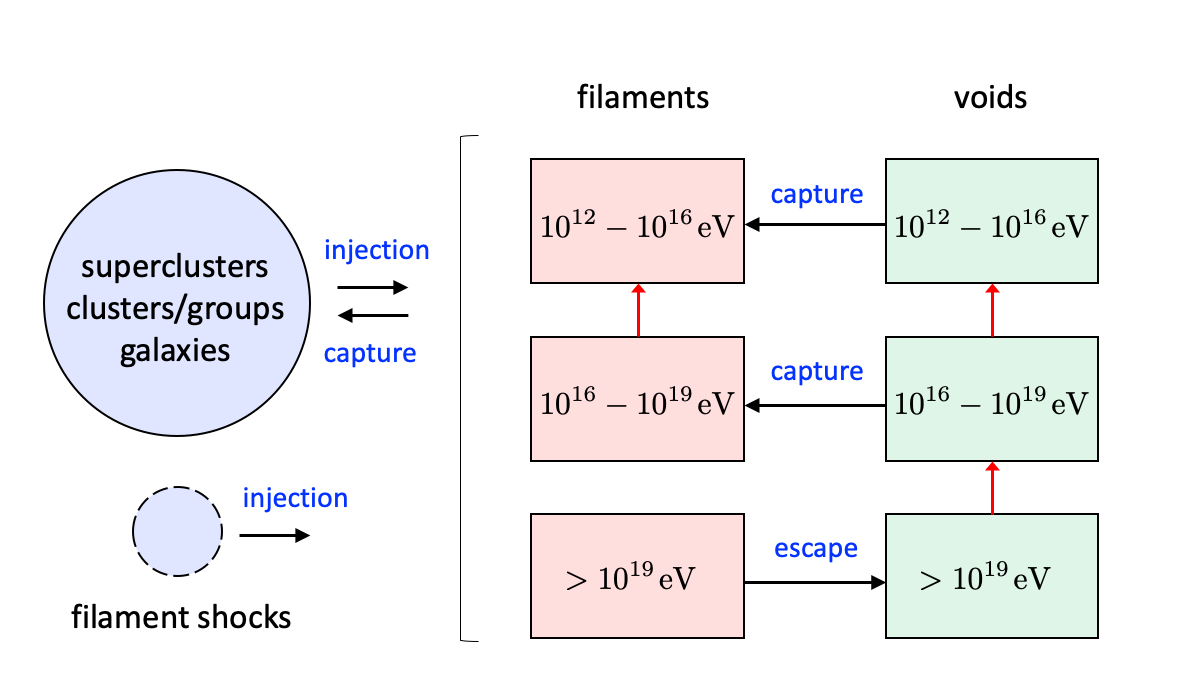} 
\end{adjustwidth}
\caption{A schematic illustration 
  of the transfer of cosmic ray protons 
  in filament and void environments.  
 The protons are sorted into three broad groups of energy 
   $>10^{19}\;\!{\rm eV}$, 
     $10^{16}-10^{19}\;\!{\rm eV}$, 
     and   $10^{12}-10^{16}\;\!{\rm eV}$.  
  The horizontal arrows denote 
   the transfer of the particles 
   in the context of relocation 
   from one astrophysical system 
   to another. 
  The vertical red arrows denote 
  the conversion of cosmic ray protons  
   from a higher-energy group 
   to a lower-energy group 
   through p$\gamma$ processes. }  
\label{fig:chart}
\end{figure}

The fact that protons with energies 
  of $10^{18}\;\!{\rm eV}$
  can be magnetically confined in filaments,  
  and that protons of energy below $10^{16}\;\!{\rm eV}$ 
  do not lose energy rapidly 
   through p$\gamma$ processes produces 
   some interesting consequences 
  in filament environments.  
We illustrate 
  schematically the transfer of cosmic rays 
  in filament environments 
  in Fig.~\ref{fig:chart}. 
From this illustration, 
 we can construct a mathematical model 
 to determine the populations 
 of cosmic ray protons in the energy range 
 $10^{12}-10^{16}\;\!{\rm eV}$ 
 within filaments  
 and voids. 

 Without loss of generality, 
  we consider an idealistic model as an illustration.  
It is expressed mathematically 
  in terms of two coupled 
  first-order differential equations: 
\begin{align}  
& \frac{{\rm d}\;\! n_{\rm fil}}{{\rm d}\;\!t}
   =  - \alpha_{\rm a}\;\! n_{\rm fil} 
    + \alpha_b n_{\rm voi} 
     + j_{\rm inj} + j_{\rm con,a} \ ; 
\label{eq:transfer_1}\\ 
& \frac{{\rm d}\;\! n_{\rm voi}}{{\rm d}\;\!t} 
   =  - \alpha_{\rm b}\;\! n_{\rm voi} + j_{\rm cov,b} 
    \ ,      
\label{eq:transfer_2}
\end{align} 
  where $n_{\rm fil}$ and $n_{\rm voi}$ are 
  the number densities of cosmic ray protons 
  in filaments and voids, respectively. 
The coefficients $\alpha_{\rm a}$ 
 and $\alpha_{\rm b}$ 
 represent the rate cosmic ray protons 
 are channelled out of filaments, and the rate 
 cosmic ray protons from voids are 
 captured by filaments. 
For filaments,  
  the rate of injection of cosmic ray protons 
  into the $10^{12}-10^{16}\;\!{\rm eV}$ energy range  
  is specified by 
  $j_{\rm inj} + j_{\rm con,a}$. Here, $j_{\rm inj}$ accounts for  
  contributions from direct injection   
  by galaxies (and galaxy groups), clusters, 
  superclusters, 
  and filament shocks (if present). 
  $j_{\rm con,a}$ is the contribution from 
protons at higher energies that are converted 
  into the $10^{12}-10^{16}\;\!{\rm eV}$ energy range 
  by p$\gamma$ processes in filaments. 
In voids, this conversion process is specified 
  as $j_{\rm con,b}$. 
Therefore, 
  this conversion process 
  tends to dominate cosmic ray injection into 
  the $10^{12}-10^{16}\;\!{\rm eV}$ energy range. 

 %
%
%
%

In general, 
 $j_{\rm conv,a}$ and $j_{\rm conv,b}$ are time-dependent. Their evolution is determined by 
 the cosmic ray energy spectrum, the cosmic ray population and local radiation fields. 
By contrast, $j_{\rm inj}$ is determined by   
  the star-formation and active galactic nuclei activities 
  at a given cosmological epoch. 
This implies that $\alpha_{\rm b}\ n_{\rm fil}$ 
  is also time-dependent, 
   even without consideration 
   of the structural evolution of filaments 
   and the cosmological expansion of voids. 
Nonetheless, 
 as the capture of cosmic ray protons in filaments 
 by lower-ordered structures 
 would not be high,  
 and as cosmic ray protons are not expected 
 to show an upturn in their spectra 
 for energies above $10^{19}\;\!{\rm eV}$, 
 it is reasonable to expect that 
 $j_{\rm inj} >  j_{\rm con,a} \gg  
  | \alpha_b n_{\rm voi} 
  - \alpha_{\rm a}\;\! n_{\rm fil}\vert$, 
 and $j_{\rm con,b} > \alpha_{\rm b}n_{\rm voi}$, 
 at least during the epochs  
 when star-formation and AGN activity peaked 
 (roughly at $z\approx 2$).  
If we keep only the dominant terms, 
  the cosmic ray transfer equations 
  are decoupled, giving a solution: 
\begin{align}  
& n_{\rm fil} \approx  
  \int_{z_{\rm AGN,0}}^0
  {\rm d}z \ 
   \left(\frac{{\rm d}z}{{\rm d}t} \right)^{-1}  \ 
   \varpi(z)  \ 
   j_{\rm int}(z) 
    \ \theta (z-z_{\rm fil,0})   
   \nonumber \\ 
  & \hspace*{2cm} + \int_{z_{\rm gal,0}}^0  
    {\rm d}z \ 
   \left(\frac{{\rm d}z}{{\rm d}t} \right)^{-1}  
    [\;\! 1 - \varpi(z) \;\!] \  
   j_{\rm int}(z) \ \theta (z-z_{\rm fil,0})  
    \ ; 
\label{eq:n_fil} \\ 
& n_{\rm voi} \approx  
   \int_{z_{\rm max}}^0
  {\rm d}z \ 
   \left(\frac{{\rm d}z}{{\rm d}t} \right)^{-1}  \ 
    \varkappa(z) \  j_{\rm int}(z) \ , 
\label{eq:n_voi}
\end{align}
  where $\theta(\cdots)$ is a Heaviside step function. 
The redshifts 
  $z_{\rm AGN,0}$,   
  $z_{\rm gal,0}$, 
  and $z_{\rm fil,0}$ 
  denote the epoch of emergence of the first AGN,   
  the first star forming galaxies,  
  and the first cosmological filaments, respectively, and 
   $z_{\rm max} 
   = {\rm Max}(z_{\rm gal,0},\;\!z_{\rm AGN,0})$. 
The ratio of relative contributions  
  of cosmic ray injection 
  by star-formation to that by AGN activities 
  at redshift $z$ is specified by 
  by $[1-\varpi(z)]/\varpi(z)$. 
The scaling for protons 
  with energies above 
  the p$\gamma$ interaction threshold 
  at redshift $z$ 
  leaking into cosmic void 
  is specified by $\varkappa(z)$.

While this set of solutions  
  does not capture   
  all of the fine details 
  of the evolution of a population of 
  cosmic ray protons 
  in the energy range $10^{12}-10^{16}\;\!{\rm eV}$, 
  it still gives us some useful insights 
  into the history of the cosmic ray content of filaments 
  and voids. 
As $({\rm d}z/{\rm d}t)$ is negative while  
 all the other terms in integrand 
 are positive, 
 $n_{\rm fil}$ and $n_{\rm voi}$ 
 would increase over time   
 until reaching a saturation level, where 
 the injection and leaking of cosmic ray protons 
 in filaments and voids fall into a steady balance.  
Filaments and voids would therefore continue 
  to accumulate cosmic ray protons in this energy range, 
  provided that their injection rate 
  is higher then their loss rate 
  through leaking into lower-ordered 
   cosmological structures, 
  or through subatomic processes  
  such as pair production.  
This implies that, 
  compared to galaxies (and groups), 
  clusters and superclusters, 
  cosmic rays in filaments and voids would have a 
    stronger component 
     of protons at energies 
    around $10^{16}\;{\rm eV}$ which are unable to cool efficiently  
    and do not undergo 
    p$\gamma$ processes.  
Filaments cannot be populated 
  by cosmic ray protons with energies 
  greatly exceeding $10^{17}\;\!{\rm eV}$, 
  because they will be depleted 
  by interactions with photons 
  in the local and cosmological radiation fields. 
They also cannot be greatly populated 
  by cosmic ray protons at energies far below 
  $10^{12}\;\!{\rm eV}$, 
  as these 
  would be magnetically confined 
  by the galaxies, clusters or superclusters 
 from where they originate, 
  and cannot migrate out into filaments and voids. 

%
\subsection{Some remarks} 
\label{subsec:remark}

%
\subsubsection{Cosmic ray energy density 
  and energy partition with magnetic fields} 
\label{subsubsec:partition}

The large population of 
  cosmic rays with energies $10^{12}-10^{16}\;\!{\rm eV}$ in filaments and voids 
  result from previous 
  star-forming and AGN activities 
  in the Universe.  Their presence 
   has some consequences  
  for how we interpret observations  
  of the micro-physics in filaments and voids. 
Firstly, 
  the spectrum of cosmic rays observed on Earth 
  and the spectrum of Galactic cosmic rays 
  are not representative 
  the spectrum in filaments and voids. 
Secondly, 
  without reliable information 
  about the cosmic ray spectrum  
  in intergalactic space, 
  caution is needed when deriving 
    certain properties of filaments and voids 
   based on the assumed number density 
   and energy spectra of cosmic rays, 
   for example when estimating filament and void magnetism.  

In the two-component deposition 
  of the filament magnetic field,   
  the two orthogonal sub-components 
  of the large-scale ordered field component 
  are ${\boldsymbol B}_{\rm g} = {\boldsymbol B}_\parallel 
  + {\boldsymbol B}_\phi$, 
  where ${\boldsymbol B}_\parallel$ 
  is aligned with filament 
  and ${\boldsymbol B_\phi}$ 
  is the toroidal field component  
  perpendicular to ${\boldsymbol B}_\parallel$. 
Including also the small-scale disordered field component, 
  the magnetic energy density 
  in the IGM within a filament would be 
\begin{align}
\langle \epsilon_{B} \rangle & = \frac{1}{8\pi}  
\left(
 \langle {B_\parallel}^2 \rangle
 + \langle {B_\phi}^2 \rangle
 + \langle {B_{\rm s}}^2\rangle  \right) \     
\end{align} 
  (assuming that the large-scale and the disordered components 
    are linearly independent).   
Unless there is strong co-evolution 
  of $\langle \epsilon_{\rm B} \rangle$ 
  and $n_{\rm fil}$,  
  it is not obvious how a relation can be established 
   between $\langle \epsilon_{B} \rangle$ 
   and $n_{\rm fil}\;\! \langle {\mathcal E}_{\rm p} \rangle$,  
    where $\langle {\mathcal E}_{\rm p} \rangle$ 
    is the energy content per cosmic ray particle  
    in the energy range $10^{12}-10^{16}\;\!{\rm eV}$. 
Also, whether or not energy equipartition 
  $\langle \epsilon_{B} \rangle = 
  n_{\rm fil}\;\! \langle {\mathcal E}_{\rm p} \rangle$ 
  can be attained uniformly 
  over a length-scale comparable 
  to the linear sizes of filaments or voids 
  is beyond our current knowledge. 

%
\subsubsection{Cosmic ray transfer on cosmological scales}
\label{subsubsec:CR_transfer} 

Treating the magnetised IGM 
  as a uniform static medium 
  gives tremendous simplification 
  when  mathematically formulating 
 cosmic ray transfer. 
One advantage is that 
  we may classify different transport regimes 
  based on the statistical properties of the IGM, 
  and this makes solving the  
   transfer equations tractable. 
However, the disadvantage with this approach  
  is that it fails to 
  account for the complexities 
  in the cosmic ray transfer process, 
  such as expanding volume, 
  multi-scale (but non-turbulent) structures. 
As shown in the case studies 
  of the journey and fate of 
  individual cosmic ray protons at different energies, 
 the regime-based solution schemes 
  using the average statistical properties the 
  of IGM 
  do not immediately 
  give a reliable description 
  of the cosmic ray properties 
  in filaments and voids.  
It is also not easy 
  to incorporate  
  the conversion of particles 
  and the sudden jump 
  in the characteristic scales 
  involved
  when cosmic rays 
  are transferred from 
  one cosmological component 
  to another, 
  in the coexistence of 
  deterministic 
  and stochastic chance encounters.  
  
The heterogeneity of the medium 
  over which cosmic rays propagate,   
  and the capture of cosmic rays  
  by strongly magnetised substructures 
  are subjects of concern 
  on galactic scales. 
Studies have been conducted 
   on quantifying how cosmic rays 
   are transferred in media with 
   intermittent patchy structures 
   \citep[see, e.g. ][]{Bustard2021ApJ, Butsky2024MNRAS}, 
   and in the presence 
   of random magnetic traps  
   \citep[][]{Tharakkal2023PhRvE}. 
The challenges of cosmic ray transport  
  on cosmological scales 
  share some similarity 
  with those 
  within the ISM of galaxies, 
  but there are also additional layers 
  of complexity 
  inherent from 
  different nature between 
  cosmological environments 
  and sub-galactic environments.  
The first is  
  how the presence of interfaces between systems 
  (see Sec.~\ref{subsec:interface})
  can play a role\footnote{This is analogous to determining the electric and magnetic 
  fields in propagating electromagnetic 
  waves across two dielectric media 
  with different refractive indices. 
  The matching of the field components 
    at the boundary is essential 
    to obtain a correct description of 
    the transmission and reflection of 
    electromagnetic waves 
    when crossing the interface 
    between the two media.}.  
The second is the operation of 
  the sieve mechanism 
  in filament environments, 
  in particular, how the conversion of particles 
  will alter the transmission 
  of cosmic rays across the interfaces. 

The traditional mathematical formulations 
   for diffusion and scattering 
   are insufficient to describe 
   cosmic ray transfer in filament environments.  
Mathematical formulations 
  of this kind 
  generally give solutions  
   in terms of a Brownian random walk,  
   or a modified version of it 
   \citep[see e.g.][]{LeVot2017PhRvE}. 
The presence of a long tail 
  in the free-path distribution 
  when cosmic rays propagate  
  through the vast intergalactic space 
  cast by an interweaving web of filaments  
  \citep[see e.g.][]{Libeskind2018MNRAS} 
  implies that 
   the underlying process 
   is Levy flight 
   \citep[see e.g.][]{Dubkov2008IJBC,Humphries2014JThBi}  
   rather than Brownian motion. 
With appropriate modifications, 
  the Levy flight formulation 
  would be able to handle complexities 
  arising from interface-induced 
  barrier crossing \citep[cf.][]{Dubkov2016EL}  
  and chance conversion/elimination of particles  
  \citep[][]{Garbaczewski2017PhRvE}. 
Constructing and solving cosmic ray transfer equations  
  with a Levy flight formulation 
  is beyond the scope of this study~\cite[see also][]{Zhang2023ApJ},  
  and we shall present our work on this  
  in a series of future papers. 

%
\subsubsection{The cosmic rays that will never reach us} 
\label{subsubsec:never-reach_us}  

Our analysis has demonstrated  
 the presence 
  of a large population of 
  cosmic ray protons in filaments in the energy range 
  $10^{12}-10^{16}\;\!{\rm eV}$. 
In a filament with a 
  characteristic magnetic field   
  $\mathcal B \sim 10\;\!{\rm nG}$, 
  the synchrotron cooling timescales 
  are about $2.2\;\!{\rm Gyr}$ 
   and $220\;\!{\rm Gyr}$ 
  for protons with energies 
  $10^{12}\;\!{\rm eV}$ 
  and  $10^{16}\;\!{\rm eV}$, 
  respectively,   
  adopting the  expression for 
  proton synchrotron cooling time 
  in \cite{Aharonian2002MNRAS}.  
Thus, there would be a pile-up 
 of energetic protons 
 above $10^{12}\;\!{\rm eV}$ 
 over time. 
In a void  
  with a magnetic field of 
 $10^{-6}\;\!{\rm nG}$, 
  the synchrotron cooling time 
  of these protons is  
  many orders greater than 
  the Hubble time. 
These wandering protons 
  in comic voids with energies 
  below the p$\gamma$ threshold 
  therefore become fossilised 
  after their last hadronic 
    interaction. 
   
Given the large combined volume occupied 
  by voids and filaments 
  (excluding galaxies, clusters and 
  superclusters), 
  the total number of cosmic ray protons 
  and the amount of energy they store
  would be substantial. 
As these hidden particles 
  have derived their energies 
  from star-formation and AGN activities,   
  they are fossil records 
  of the power generation 
  history of the Universe 
  (after the emergence 
  of the first stars, first 
  galaxies and first quasars). 

The cosmic ray spectrum 
  observed on Earth is characterised 
  by features such as 
  the ankle and knees   
  \citep[see e.g.][]{Gaisser2013FrPhy}. 
A simple interpretation 
   \citep[as illustrated schematically by][]{Alves_Batista2019FrASS} 
  is that the cosmic rays 
  are dominated 
  by three components.  
The knees at energies 
  below a few $10^{17}\;\!{\rm eV}$   
  are signatures of 
  the two lower-energy components, 
  which are of Galactic origin,   
  contributed mainly by supernova remnants.  
The ankle is caused by the transition 
  from the two lower-energy components 
  to a high-energy component.  
It is commonly attributed 
  to cosmic rays originating 
  from outside the Milky Way 
  \citep[for a review, see][]{Aloisio2012APh}.   
The extragalactic component 
  has a low-energy drop-off, 
  as cosmic rays with energies 
  below $\sim 10^{16}\;\!{\rm eV}$ 
  from outside the Milky Way   
  are expected to be strongly deflected 
  by Galactic magnetic fields.  
The observed cosmic ray spectrum 
  on Earth implies that 
  galaxies like the Milky Way 
  will censor the direct detection of   
  of the fossil cosmic ray 
  proton populations,   
  even if these protons 
  manage to leave the filaments 
  and voids they had previously resided in. 
In other words, these cosmic rays 
  will never reach us. 

The cosmic ray spectrum observed on Earth is 
 a consequence of a dynamical equilibrium 
 of cosmic ray transfer from Galactic and 
 extragalactic origins. 
The spectral properties reflect 
  the ``now'' situation of 
  the cosmic ray properties 
  around the Earth, 
  and the  synchronisation 
  of  the evolution of 
  the cosmic ray spectrum  
  with cosmic ray production,
  transport, conversion and destruction 
  in the nearby Universe. 
The spectral properties of 
  the cosmic ray population 
  of energies $10^{12}-10^{16}\;\!{\rm eV}$ 
  in filaments and voids 
  are, by contrast, 
  the result of a dynamical equilibrium. 
This cosmic ray population is evolving 
  with the Universe.  
Their cumulative nature 
  implies that 
  they retain memory 
  of past events (see Eq.~\ref{eq:n_fil} 
  and \ref{eq:n_voi}), 
  such as the epochs 
  when star-forming and AGN activities 
  in the Universe peaked,  
  and the continual reconfiguration 
  of the cosmic web 
  woven by large-scale filaments.

\section{Conclusions}  

In this work, we have 
demonstrated the importance of cosmological filaments in 
the transport and entrapment of energetic hadronic cosmic rays. 
We have found that the interplay of energy-dependent particle transport 
and hadronic interaction processes 
produces a 
range of evolutionary paths for the cosmic rays 
depending on their energy and location of origin. 
In particular, two regimes emerge in which cosmic rays engage differently with filament ecosystems. 
At low energies, they undergo transport, but are below the 
threshold energy for pion production processes. Without any 
other efficient cooling process available for them, 
the evolution of these cosmic rays is regulated by 
the magnetic configuration of filament ecosystems, 
with their energy becoming `frozen'.

At higher energies, above the pion production threshold,  
cosmic rays degrade in energy via interactions with cosmological and local radiation fields as they propagate. 
These interacting cosmic rays 
can generally escape from 
magnetized structures embedded within 
filaments, but only some may be able to diffuse out of the large filaments and into cosmic voids.  
They lose energy quickly through pion production, and soon fall below the interaction threshold energy. They then 
join the fossil population of non-interacting cosmic rays in voids and filaments. 

The exact fate of cosmic rays depends on their location of birth. If originating from a cluster or galaxy, only the most energetic cosmic rays have a chance to diffuse out. Lower energy cosmic rays are trapped in their cluster or galaxy of origin.  If originating from a filament, 
only the highest energy cosmic rays can escape. 
Those below 10$^{18}\;\!{\rm eV}$ are confined 
by the filament to scatter and diffuse inside it. The filament then 
operates as a cosmic ray highway, channeling the entrapped particles along it. Those cosmic rays which do escape 
 lose energy in the void by pion production until they fall below the interaction threshold. Unless these fugitive cosmic rays are captured by a filament while they shed their energy, they are left to wander in the void. Unable to cool efficiently, they form a relic ocean of void-filling cosmic rays. They can 
 only leave the void by colliding 
  directly with a filament or one of its gravitationally bound substructures where the stronger magnetic fields can capture them.  

Overall, our findings point towards a build-up of a cosmic ray ocean in the $10^{12}-10^{16}\;\!{\rm eV}$  
energy range within voids and filaments. At higher energies, cosmic ray populations are depleted by interactions with photons in local and cosmological radiation fields. At lower energies, cosmic rays can be magnetically confined by the galaxies, clusters or superclusters from where they originate, and cannot migrate out into filaments and voids. 
This ocean of relic cosmic rays in filaments and voids develops to harbour a substantial energy component in the Universe. Their exact spectral properties and cumulative evolution become 
 a long-lived `fossil' population 
 that 
 records the power generation history of the Universe and the evolution of the cosmic web.   
Yet, as they would be 
 strongly deflected by the magnetic fields of galaxies like the Milky Way, 
 this vast population of relic cosmic rays will never reach us, even if they 
are able to escape from their structure of origin.

\authorcontributions{
All authors contributed to the scientific content of this article. 
Q.H. cross-checked all the calculations thoroughly. 
K.W. and E.R.O. prepared most of the manuscript on behalf 
  of the authors. 
}

\acknowledgments{K.W. and Q. H. 
  acknowledge support from the UCL Cosmoparticle Initiative. 
Q.H. is supported by a UCL Overseas Research Scholarship. 
Q.H. and L.L. are supported 
  by UK STFC Research Studentship. 
E.R.O. is an overseas researcher 
  supported by a Postdoctoral Fellowship 
  of the Japan Society for the Promotion of Science 
  (JSPS KAKENHI Grant Number JP22F22327).  
Y.I. is supported by an 
  NAOJ ALMA Scientific Research Grant Number 2021-17A, 
  JSPS KAKENHI Grant Numbers JP18H05458, JP19K14772, and JP22K18277, 
  and the World Premier International Research Center Initiative 
  (WPI), MEXT, Japan. 
This work made use of the NASA Astrophysics Data System (ADS). }

\conflictsofinterest{The authors declare no conflict of interest.} 

\begin{table}[h] 
\vspace*{0.25cm}
\caption{List of abbreviations in the main text.}
\begin{tabular}{lll}
\hline
Abbreviation & Definition \\
\hline
AGN& Active galactic nuclei/nucleus \\
CGM& Circumgalactic medium\\
CMB& Cosmic microwave background\\
GZK& Greisen-Zatespin-Kuzmin\\
ICM& Intra-cluster medium\\
IGM& Intergalactic medium\\
IR& Infra-red\\
ISM& Interstellar medium\\
$\Lambda$CDM& Lambda Cold Dark Matter\\
RM& Rotation measure\\
UHE& Ultra high-energy \\
WHIM& Warm-hot intergalactic media/medium\\
\hline
\end{tabular} 
\vspace*{0.15cm} \\ 
\label{tab:abb}
\end{table}

\begin{table}[h] 
\vspace*{0.25cm}
\caption{List of abbreviations in Appendix A.}
\begin{tabular}{lll}
\hline
Abbreviation & Definition \\
\hline
EBL& Extragalactic background light\\
FUV& Far-ultraviolet\\
HCG& Hickson Compact Group\\
HyLIRG& Hyper-luminous infra-red galaxy\\
ISRF& Interstellar radiation field \\
IGrM& Intra-group medium\\
\hline
\end{tabular} 
\vspace*{0.15cm} \\ 
\label{tab:abb_app}
\end{table}





\appendixtitles{no} 

\appendixstart
\appendix

\section[\appendixname~\thesection]{}
\label{app:AAAAA}  

\begin{table}[H] 
\caption{Summary of the parameters adopted in our calculations for the hadronic pp and p$\gamma$ 
   path lengths shown in 
Fig.~\ref{fig:hadronic_ints}, 
  for filament and void conditions. 
  In all cases, p$\gamma$ interactions with the CMB at the specified redshift are included in our calculations.} 
\begin{adjustwidth}{-\extralength}{0cm}
\newcolumntype{C}{>{\centering\arraybackslash}X}
\begin{tabularx}{\fulllength}{lCCCCC}
\toprule
\multirow[m]{2}{*}{\textbf{Environment}}	& \multirow{2}{*}{\textbf{Redshift}} & \multicolumn{2}{c}{\textbf{Radiation Energy density [${\rm eV\;\!cm}^{-3}$]}} &  \textbf{Gas density} & \textbf{Size}$^{(h)}$ \\
& & Starlight & Dust & \textbf{[g cm$^{-3}$]} & \textbf{[Mpc]} \\
\midrule
\midrule
\multirow{3}{*}{Central filament$^{(a)}$} & 0 & 3.7 & 5.2 & 4.0\;\!e-29 & 0.30 \\  
& 2 & 28 & 42 & 3.6\;\!e-28 & 0.20 \\ 
& 7 & 3.4 & 4.7 & 2.0\;\!e-26 & <\;\!0.050 \\
\midrule
\multirow{3}{*}{Filament outskirts$^{(a)}$} & 0 & 0.10 & 0.14 & 1.4\;\!e-30 & 2.0 \\   
& 2 & 0.076 & 1.1 & 1.3\;\!e-29 & 2.5 \\  
& 7 & 0.091 & 0.13 & 7.0\;\!e-28 & >\;\!2.8 \\
\midrule
\multirow{3}{*}{Void$^{(b)}$} & 0 & 0.022 & 0.032 
& 8.0\;\!e-32 & 7.9 \\ 
& 2 & 0.17 & 0.25 & 1.4\;\!e-30 & 6.7 \\  
& 7 & 0.021 & 0.028 & 7.6\;\!e-29 & 6.0 \\
\midrule
\multirow{3}{*}{Average IGM$^{(c)}$} & 0 & 0.024 & 0.035 
& 4.0\;\!e-31 & -- \\
& 2 & 0.19 & 0.29 & 3.6\;\!e-30 & -- \\
& 7 & 0.023 & 0.031 & 2.0\;\!e-28 & -- \\
\midrule 
Starburst galaxy$^{(d)}$ & 0 & 670 & 310 & 1.7\;\!e-20 & 0.0010 \\
\midrule 
\multirow{1}{*}{CGM$^{(e)}$} & 0 & 0.24 & 0.34 & 1.0\;\!e-28 & 0.10 \\
\midrule 
\multirow{1}{*}{Intra-group medium$^{(f)}$} & 0 & 0.22 & 0.31 &  4.4\;\!e-28 & 0.12 \\
\midrule 
\multirow{1}{*}{Intra-cluster medium$^{(g)}$} & 0 & 0.21 & 0.28 & 1.1\;\!e-27 & 1.9 \\
\bottomrule
\end{tabularx}
\label{tab1:hadr_int_params}
\end{adjustwidth}
\end{table}

\begin{adjustwidth}{-\extralength}{0cm}\footnotesize{
\vspace{-0.4cm} 
\textbf{Notes:} \\
\vspace{-0.2cm} \\
$^{(a)}$ Filament gas density estimates 
  are based on the structural classification 
  proposed by Ref.~\cite{Dolag2005JCAP}, 
   where central filaments have an overdensity of 100, 
   and filament outskirts have an overdensity of 3.5 
   compared to the average background IGM. 
Central densities are typically 5-15 times 
  the critical density of the Universe at all redshifts.  
The energy densities of filament radiation fields follow the total stellar and dust contributions to the extragalactic background light (EBL) at $z=0$ in Ref.~\cite{Dermer2009herb}. 
These are modeled 
 as modified black-bodies, 
 with characteristic 
 temperatures of 7,100 K (starlight) and 62 K (dust), 
 following the dominant components 
 of the EBL. 
Radiation energy densities 
  in filaments at higher redshifts 
  are obtained 
  by scaling the $z=0$ EBL values 
  with the cosmic star-formation rate density \cite{Madau2014ARAA}, as obtained from FUV and IR data for the stellar and dust EBL components, respectively. Within filaments, radiation fields are scaled by the excess stellar density for long filaments in Ref.~\cite{Galarraga-Espinosa2022A&A}, 
corresponding to an increase by a factor of 150 in central filaments, and by 4 in filament outskirts.  \\
$^{(b)}$ Void densities $\rho_{\rm voi}$ are estimated from typical density contrasts of $\rho_{\rm voi}/\rho_{\rm B}\sim 0.20$ at $z=0$ and 0.38 at $z=2$
compared to the background average IGM density $\rho_{\rm B}$~\cite{Ricciardelli2013MNRAS}. 
An estimate for the void density contrast is unavailable at $z=7$. Given that void density contrasts are weaker at higher redshifts, we apply the $z=2$ contrast to ensure our path length calculations are conservative. Our estimated void densities are a factor of $\sim$ 10-100 lower than the critical density at all redshifts.  
The reduction of the EBL due to the presence of a void is not substantial. 
Following Ref.~\cite{Abdalla2017ApJ}, we assume 
 a 10 per cent reduction compared to the average EBL energy density. 
This is conservative, and corresponds to voids of sizes 
100 $h^{-1}$ Mpc which have the lowest EBL levels. The $\Lambda$CDM model of Ref.~\cite{Adermann2018MNRAS} is adopted for the redshift evolution of cosmic void sizes. \\
$^{(c)}$ The comoving cosmic mean baryon density 
  reported by Ref.~\cite{Walter2020ApJ} 
  is adopted as the average IGM density at $z=0$. 
alues at higher redshifts are scaled from the $z=0$ density by cosmological volume.  \\ 
$^{(d)}$ Parameters are informed by conditions in the nearby Hyper Luminous IR starburst galaxy (HyLIRG) IRAS F14537+1950, where the interstellar radiation field (ISRF) intensity is scaled by bolometric luminosity compared to the Galaxy, 
from~\cite{Rowan-Robonson2000MNRAS}. F14537+1950 has a redshift of $z=0.64$ with no indication of AGN activity 
 \cite{Rowan-Robonson2000MNRAS}. 
This is chosen as an extreme example 
  of a low-redshift starburst galaxy. 
To allow for direct comparison with the other structures, our path length calculations assume $z=0$, rather than the measured redshift of this galaxy. The external EBL contribution is negligible. \\ 
$^{(e)}$ Gas density estimated from the mean  CAMELS-IllustrisTNG profiles in Ref.~\cite{Moser2022ApJ} at a radius of 0.1 Mpc for galaxies of halo mass between $10^{12}\;\!{\rm M}_{\odot}$ and $10^{12.3}\;\!{\rm M}_{\odot}$. At this radius, feedback effects do not have a large impact on the density profile. Radiation field energy densities are estimated by scaling the dust emission from the ISM of the starburst galaxy IRAS F14537+1950 by the square of the relative sizes of the systems. This assumes that most stellar radiation in a starburst galaxy is re-radiated in the IR band by dust. 
The starlight is then scaled from this by the relative energy density ratio of the EBL. An additional contribution from the average external IGM EBL is included in the CGM radiation field. \\
$^{(f)}$ Properties of groups of galaxies show broad variation.
We adopt the average properties of the sample of galaxy groups from Ref.~\cite{Lagana2013A&A} 
(see $r_{500}$ values in their Table 1a) as a representative example of the density of the intra-group medium (IGrM) (for a review of properties of the IGrM, 
see Ref.~\cite{Oppenheimer2021Univ}, 
and characteristic physical/observable properties of galaxy compact groups, see Ref.~\cite{Bitsakis2011A&A}). 
For the intra-group light (IGrL), we estimate the dust contribution using the total IR luminosity of the Hickson Compact Group (HCG) 40 group members in~\cite{Bitsakis2011A&A} (see their Table 4) and separations (their Table 6), adopting a 5 member configuration. The stellar contribution is then scaled from the dust contribution according to the ratio of starlight to dust luminosities reported for the EBL at $z=0$. An additional contribution from the average IGM EBL is included in the IGrL radiation field, which permeates the system. \\
$^{(g)}$ The density of the intra-cluster medium 
is based on the average of the sample of clusters presented in Ref.~\cite{Lagana2013A&A} (see $r_{500}$ values in their Table 1b). 
This value corresponds to $\sim$130 times the critical density of the Universe at $z=0$. 
When accounting for the uncertainties in the sample, 
this overdensity is comparable to 
that typically adopted for galaxy clusters, 
where a value of $\sim$ 200 has been shown by N-body simulations to correspond to the virialized region of a cluster's dark matter halo~\cite[see][]{Cole1996MNRAS}. 
Radiation energy densities are considered to be double those of the filament outskirts, roughly following the increase in radiation fields shown in Ref.~\cite{Longobardi2020A&A}, if estimating characteristic values from the Virgo cluster.  
This is conservative to ensure that cosmic ray interaction rates are not overstated in our calculations. 
Note that the adopted values are lower than in central filaments, as they are an average value over the cluster. In central cluster regions, radiation fields may be comparable to or could even exceed those of the central regions of filaments. The contribution from the external EBL is intrinsically included by our approach. \\
$^{(h)}$ Characteristic proper widths of filaments are obtained from Ref.~\cite{Galarraga-Espinosa2023arXiv}, which provides density profiles up to $z=4$ 
showing the collapse and contraction of the filament profile. At $z=7$, the result at $z=4$ is taken as a limit. 
The central filament is considered to be the region where gas densities in Ref.~\cite{Galarraga-Espinosa2023arXiv} are 100 times the background IGM average density, while the outskirts are considered to be 3.5 times higher than the background IGM. 
Void sizes are estimated 
by invoking spherical morphologies to convert  
characteristic proper volumes reported for the $\Lambda$CDM model of Ref.~\cite{Adermann2018MNRAS} to proper 
void diameters, assuming a Hubble parameter of $h=0.7$. \\}
\end{adjustwidth}


\reftitle{References}  
\bibliography{filament.bib}

\PublishersNote{}
\end{document}